\documentclass[twocolumn,aps,superscriptaddress,multicol,amsmath,amssymb]{revtex4-1}
\usepackage{amssymb}
\usepackage{amsmath}
\usepackage{graphicx}
\usepackage{sidecap}
\usepackage{epstopdf}
\usepackage{bm}%
\usepackage{ulem}%
\usepackage{gensymb}
\usepackage{soul}
\usepackage{sidecap}
\usepackage{bbm}
\usepackage{times}
\usepackage[utf8x]{inputenc}

\newcommand{\bra}[1]{\ensuremath{\langle #1 |}}
\newcommand{\ket}[1]{\ensuremath{| #1 \rangle}}

\newcommand{\be}{\begin{equation}}
\newcommand{\ee}{\end{equation}}
\newcommand{\bk}{{{\bf{k}}}}
\newcommand{\bq}{{{\bf{q}}}}

\newcommand{\br}{{{\bf{r}}}}

\newcommand{\bg}{{{\bf{g}}}}

\newcommand{\bR}{{{\bf{R}}}}

\newcommand{\bea}{\begin{eqnarray}}
\newcommand{\eea}{\end{eqnarray}}

\newcommand{\bd}{\begin{displaymath}}
\newcommand{\ed}{\end{displaymath}}
\newcommand{\ba}{\begin{array}}
\newcommand{\ea}{\end{array}}
\newcommand{\bi}{\begin{itemize}}
\newcommand{\ei}{\end{itemize}}
\newcommand{\bc}{\begin{center}}
\newcommand{\ec}{\end{center}}
\newcommand{\bfl}{\begin{flushleft}}
\newcommand{\efl}{\end{flushleft}}
\newcommand{\bfr}{\begin{flushright}}
\newcommand{\efr}{\end{flushright}}

\newcommand{\mi}{\rm i}

\newcommand{\bl}{\begin{aligned}}
\newcommand{\el}{\end{aligned}}

\def\ket#1{\left\vert #1 \right\rangle}
\def\br{{\bf r}}\def\bR{{\bf R}}
\def\bk{{\bf k}} \def\bq{{\bf q}}  
\def\bg{{\bf g}}  \def\bd{{\bf d}}

\def\6{\partial}

\def\bra{\langle}
\def\ket{\rangle}
\def\={\!\!\!&=&\!\!\!}
\def\+{\!\!\!&&\!\!\!+~}
\def\-{\!\!\!&&\!\!\!-~}

 \newcommand\redsout{\bgroup\markoverwith{\textcolor{red}{\rule[0.5ex]{2pt}{0.4pt}}}\ULon}
\newcommand\bluesout{\bgroup\markoverwith{\textcolor{blue}{\rule[0.5ex]{2pt}{0.4pt}}}\ULon}
\newcommand\redout{\bgroup\markoverwith{\textcolor{red}{\rule[.5ex]{2pt}{0.4pt}}}\ULon}

\usepackage[colorlinks=true,citecolor=blue]{hyperref}
\hypersetup{colorlinks=true,citecolor=blue,linkcolor=red,urlcolor=blue}

\graphicspath{{.}{./figs/}}
\begin{document}
\title{Spin and charge order in doped spin-orbit coupled  Mott insulators} 

\author{Mehdi Biderang} 
\email{mehdi.biderang@umanitoba.ca}
\affiliation{Department of Physics and Astronomy, University of Manitoba, Winnipeg R3T 2N2, Canada}
\affiliation{Manitoba Quantum Institute, University of Manitoba, Winnipeg R3T 2N2, Canada}
\author{Alireza Akbari}
\email{akbari@postech.ac.kr}
\affiliation{Max Planck Institute for the Chemical Physics of Solids, D-01187 Dresden, Germany}
\affiliation{Max Planck POSTECH Center for Complex Phase Materials,and Department of Physics, POSTECH, Pohang, Gyeongbuk 790-784, Korea}
\author{Jesko Sirker}
\email{sirker@physics.umanitoba.ca}
\affiliation{Department of Physics and Astronomy, University of Manitoba, Winnipeg R3T 2N2, Canada}
\affiliation{Manitoba Quantum Institute, University of Manitoba, Winnipeg R3T 2N2, Canada}
\date{\today}
%
\begin{abstract}
We study a two-dimensional single band Hubbard Hamiltonian with
antisymmetric spin-orbit coupling. We argue that this is the minimal
model to understand the electronic properties of locally
non-centrosymmetric
 transition-metal (TM) oxides 
 such as
Sr$_2$IrO$_4$. Based on exact diagonalizations of small clusters and
the random phase approximation, we investigate the correlation effects
on charge and magnetic order as a function of doping and of the TM-oxygen-TM bond angle $\theta$. 
For small doping and
$\theta\lesssim 15\degree$ we find dominant commensurate in-plane
antiferromagnetic fluctuations while ferromagnetic fluctuations
dominate for $\theta\gtrsim 25\degree$. Moderately strong
nearest-neighbor Hubbard interactions can also stabilize a charge
density wave order. Furthermore, we compare the dispersion of magnetic
excitations for the hole-doped case to resonant inelastic X-ray
scattering data and find good qualitative agreement.
\end{abstract}
%
\maketitle
%
\section{Introduction}
Solid state systems with strong electron correlations show an
intriguing number of quantum many-body phenomena including
high-temperature superconductivity, spin-liquid phases, colossal
magnetoresistance, and multiferroic behavior. They can also host
exotic quasiparticles such as Majorana and Weyl fermions
\cite{Dagotto_Science_1996,Nagaosa_Science_2000,Dagotto_Science_2005,Wen_Rev_Mod_Phys_2006,Buchner_NatPhys_2009,Scalapino_Rev_Mod_Phys_2012,Taylor_PRL_2012,BJ_Kim_NatPhys_2015,Zhou_PRL_2016,Majumdar_PRL_2018,Kitagawa_Nature_2018}. 
Another recurring theme in contemporary condensed matter physics is
the emergence of various types of charge and spin order in strongly
interacting systems
\cite{BJ_Kim_Annual_2019,Weng_PhysRevRes_2020}. Interesting ordering
phenomena were experimentally observed in many rare-earth and
transition metal oxide (TMO) compounds, for example,
La$^{}_{1.6}$Nd$^{}_{0.4}$Sr$^{}_x$CuO$^{}_4$,
YBa$^{}_2$Cu$^{}_3$O$^{}_{6+x}$,
Bi$^{}_2$Sr$^{}_2$Cu$^{}_2$O$^{}_{8+x}$,
La$^{}_{1.5}$Sr$^{}_{0.5}$NiO$^{}_4$, Na$^{}_x$CoO$^{}_2$, and
Sr$^{}_2$IrO$^{}_4$ some of which also exhibit high temperature
superconductivity~\cite{Uchida_Nature_1995,Dogan_PRL_2002,Kajimoto_PRB_2003,Kapitolnic_PRB_2003,Wilson_Nat_Comm_2018}. Strong
electron interactions make TMOs a particularly promising class of
materials to find novel exotic
phases~\cite{Dagotto_Science_2005,Scalapino_Rev_Mod_Phys_2012}.

In 5d TMOs, the presence of crystal fields, spin-orbit couplings
(SOC), and strong Coulomb interactions leads to enhanced quantum
fluctuations and a competition between a variety of often exotic
ground
states~\cite{Pesin_Nature_2010,Witzcak_Krampa_Annual_2015,Kee_Annual_2016}. Among
these materials, the iridates and especially the layered perovskite
Sr$_2$IrO$_4$ has attracted a lot of attention due to its similarities
with the cuprate
superconductors~\cite{Fujiyama_PRL_2012,Wang_PRL_2011,BJKim_Science_2014,Watanabe_PRL_2013,Meng_PRL_2104,delaToore_PRL_2015,Biderang_PRB_2017,Nelson_Nat_Com_2020}. For
example, Sr$_2$IrO$_4$ (La$_2$CuO$_4$) has one hole per Ir (Cu) ion,
and shows a pseudospin-$\frac{1}{2}$ antiferromagnetic
order. Moreover, recent experiments on electron-doped Sr$_2$IrO$_4$
indicate the emergence of a pseudogap and, at low temperatures, of a
d-wave gap which strengthens the analogy with the cuprates~\cite{BJ_Kim_NatPhys_2015}. On the other hand, there are also distinct
differences. Sr$_2$IrO$_4$ has, in particular, large spin-orbit
couplings and a non-symmorphic crystal structure.

Triggered by the discovery of superconductivity in the
non-centrosymmetric (NCS) superconductor CePt$_3$Si, the role played
by antisymmetric spin-orbit coupling (ASOC) for the electronic and
topological properties of a band structure has come into
focus~\cite{Grokov_PRL_2001,Sigrist_PRL_2004,Bauer_PRL_2004,Samokhin_PRB_2004,Frigeri_PRL_2004,Fujimoto_Jpn_2007,Yanase_Jpn_2010,Bauer_Sigrist_NCS_Book_2012,Biderang_PRB_2018,Greco_PRL_2018,Greco_PRB_2020}. One
of the most intriguing features of topological non-centrosymmetric
superconductors is that they can host Majorana
fermions~\cite{Qi_RevModPhys_2011,Beenaker_RevModPhys_2015}. Locally
NCS superconductors belong globally to a centrosymmetric space group
(global inversion symmetry)~\cite{Fiscer_SIgrist_PRB_2011}, however,
as a result of randomly distributed stacking faults the inversion
symmetry is locally
broken~\cite{Yanase_PRL_2017,Yanase_PRB_2018}. Compounds belonging to
this class are, for example, Sr$_3$Ru$_2$O$_7$, Sr$_2$RhO$_4$, and
Sr$_2$IrO$_4$~\cite{Shaked_2000,SUBRAMANIAN_1994}. Here distortions in
the TMO-oxygen-TMO bonds break inversion symmetry locally and lead to
staggered ASOCs.

The crystal structure of Sr$_2$IrO$_4$ has been investigated
experimentally in a number of studies. Early neutron powder
diffraction measurements indicated that the crystal structure of
Sr$_2$IrO$_4$ belongs to the centrosymmetric space group
$I4^{}_1/acd$~\cite{Huang_JSSC_1994,Huang_PRB_1994}. More recent
studies by single-crystal neutron diffraction revealed, however, a
$I4^{}_1/a$ space group \cite{Ye_PRB_2015,Torchinsky_PRL_2015}. In
both cases, the crystal structure is globally centrosymmetric and
non-symmorphic. Since local inversion symmetry at the Ir sites is
missing, ASOC and the entanglement of various internal degrees of
freedom are expected to occur in this locally NCS
system~\cite{Yanase_PRL_2017,Yanase_PRB_2018}. Moreover, the results
of resonant inelastic X-ray scattering (RIXS) on electron-doped
Sr$_2$IrO$_4$ show that magnetic correlations persist well into the
metallic regime while the long-range magnetic order is
lost~\cite{BJ_Kim_PRL_2016,Hil_PRB_2016,Pincini_PRB_2017}. This
property is another similarity to the case of {\it hole-doped}
cuprates. I.e., there is a type of electron-hole conjugation between
the properties of the iridates and those of the cuprates
\cite{Wang_PRL_2011}. 

The problem of the interplay between spin-orbit interactions, magnetic
and charge fluctuations has been considered for a broad range of
strongly correlated electron materials, such as cuprate high-$T^{}_c$
superconductors, heavy fermion compounds, and
TMOs~\cite{Kivelson_Nature_1998}. In the particular case of the
cuprates, the role of charge and antiferromagnetic spin fluctuations
and their relation to superconductivity remains controversial. In this
paper, we will investigate the charge and magnetic properties of a
two-dimensional single band Hubbard model with antisymmetric
spin-orbit coupling describing materials such as the $5d$ layered TMO
Sr$_2$IrO$_4$. Our model is taking into account a next-nearest
neighbor hopping, which leads to an asymmetry of hole and electron
doping, as well as nearest-neighbor Hubbard interactions which can lead
to charge order. In addition, the effects of the rotation of the
IrO$^{}_6$ octahedra are included.
Our main goal is to investigate the dominant spin and charge
fluctuations across the phase diagram of this model. Apart from being
of interest in their own right, this study will also set the stage to
discuss the mechanisms for superconductivity in future studies.

The paper is organized as follows: In Sec.~\ref{Sec_Model} we
introduce the model and consider its fundamental properties in the
non-interacting case. In Sec.~\ref{Sec_ED} we then use exact
diagonalizations of small clusters to develop some understanding of
the dominant short-range magnetic fluctuations in the interacting
case. Next, we derive in Sec.~\ref{Sec_RPA} the dynamical charge and
spin susceptibilities as well as the magnon dispersions using the
random-phase approximation (RPA). The last section is devoted to a short summary and conclusion.

\begin{figure}[t]
\begin{center}
\hspace{-0.1cm}
\includegraphics[width=0.8 \linewidth]{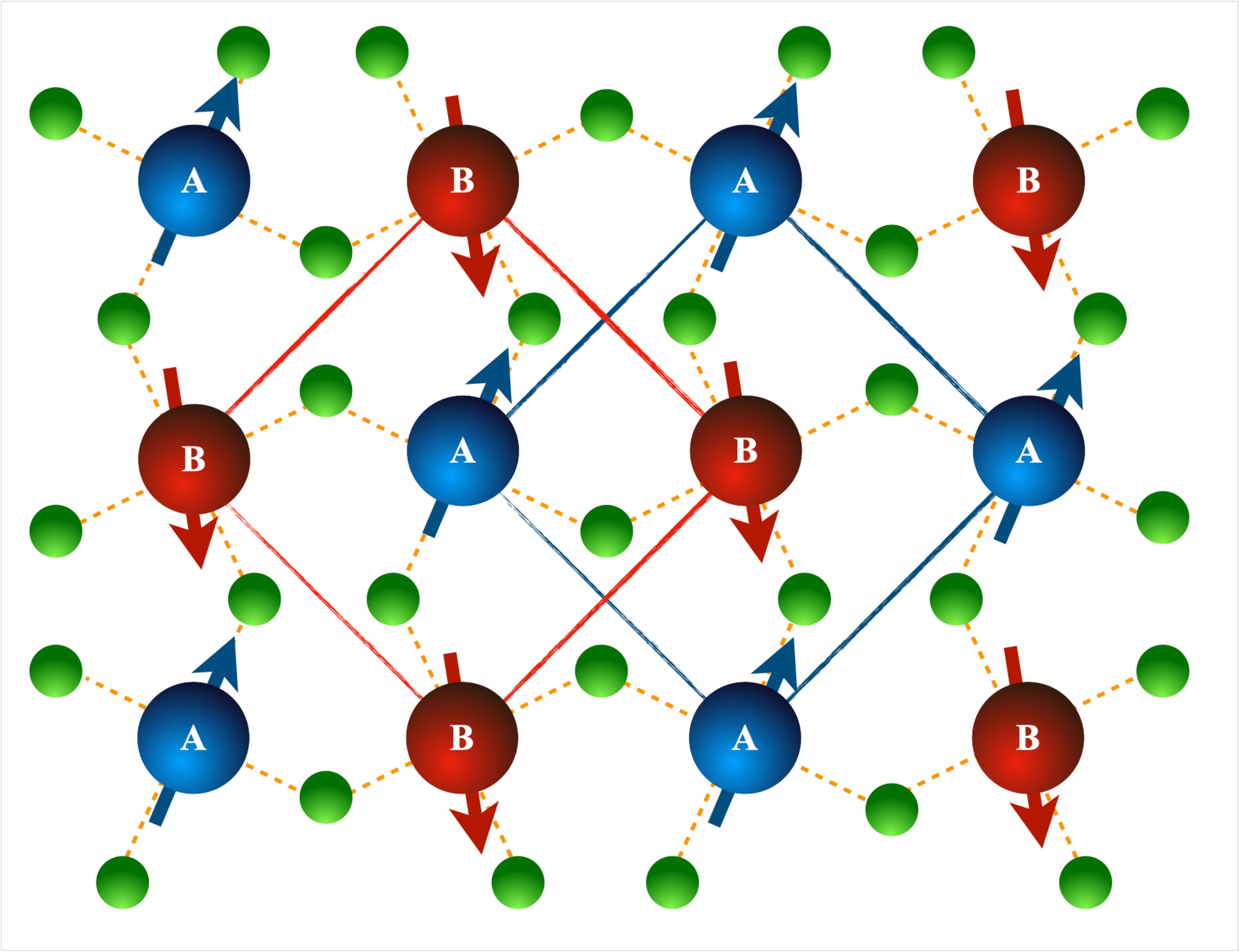} 
\end{center}
\vspace{-0.5cm}
\caption{The unit cell of the layered perovskite Sr$_2$IrO$_4$. 
The red and blue circles denote Ir atoms at different sublattices with
the letters A and B. The red and purple arrows indicate the magnetic
moments on Ir sites. The green circles are oxygen atoms. The deviation
of Ir-O-Ir bonds from $180^{\circ}$ breaks the local inversion
symmetry and generates a non-colinear antiferromagnetic order along
with a partially ferromagnetic moment.}
\label{Fig:Unit_Cell}
\end{figure}
%

\section{Description of the Model} 
\label{Sec_Model}
Sr$_2$IrO$_4$ is a layered material based on the stacking of
two-dimensional IrO$_2$ sheets, in which Ir$^{4+}$ ions form a square
lattice with two sublattices. At each lattice point, the IrO$_6$
octahedron is elongated along the c-axis and rotated around it by an
angle of $\theta \approx 11\degree$, leading to a locally broken
inversion symmetry at each sublattice. X-ray scattering and neutron
diffraction experiments detected this local non-centrosymmetricity as
a canted antiferromagnetic order
\cite{Moon_PRB_2009,Ye_PRB_2013}. Besides, measurements of the magnetic
susceptibility revealed a weak ferromagnetic moment
\cite{Hsieh_PRL_2015}. The unit cell of Sr$_2$IrO$_4$ and 
the magnetic order is shown schematically in~Fig.~\ref{Fig:Unit_Cell}.

A sufficiently large crystal field in a $5d$ TMO splits the $5d$
orbitals into a $t^{}_{2g}$ triplet and an $e^{}_g$ doublet. The
$t^{}_{2g}$ states are the low-spin ground states of the system
\cite{BJKim_PRL_2008,Senthil_prl_2011}. For strong SOC, the $t_{2g}$ 
orbitals---an effective $L=1$ systems--- is split further into fully
filled $J_{\rm eff} =3/2$ and half-filled $J_{\rm eff} = 1/2$ (upper)
states~\cite{BJKim_PRL_2008,Senthil_prl_2011}. The reduction of the
bandwidth due to SOC causes the formation of a Mott-insulating ground
state in the presence of an intermediate amount of correlation between
the electrons~\cite{Jaejun_PRB_2009}. The single-band Mott-insulating
picture for the low-energy physics of Ir oxides with $J_{\rm eff} =
1/2$ appears to be consistent with a number of experimental and
theoretical investigations. Based on this simplified single band
picture, the Hubbard-type model which we want to investigate is given
by
%
\bea
\bl
{\cal H}={\cal H}^{}_{0}+{\cal H}^{}_{\rm Int},
\label{ini_Ham}
\el
\eea
where ${\cal H}^{}_0$ and ${\cal H}^{}_{\rm Int}$ are the
non-interacting and interacting parts of the Hamiltonian,
respectively. The non-interacting part in real space can be written as
%
\begin{eqnarray}
\label{Eq:real_Ham}
{\cal H}^{}_{0} =-\mu
	&-&
	\sum_{ \bra ij\ket, \sigma}
	\Big\lbrace
	t^{}_1 
	[
	a^{\dagger}_{i\sigma} b^{}_{j\sigma}
	+
	b^{\dagger}_{i\sigma} a^{}_{j\sigma}
	]
	\\
	&+&
	\sum_{\sigma'}
	{\mi}
	t'^{}_1 
	\hat{\sigma}^z_{\sigma\sigma'}
	[
	a^{\dagger}_{i\sigma}
	b^{}_{j\sigma'}
	-
	b^{\dagger}_{i\sigma}
	a^{}_{j\sigma'}
	]
	\Big\rbrace
	\nonumber \\
	&-&
	\sum_{ \bra\!\bra ij\ket\!\ket, \sigma}
	\Big\lbrace
	t^{}_2
	[
	a^{\dagger}_{i\sigma}
	a^{}_{j\sigma}
	+
	b^{\dagger}_{i\sigma}
	b^{}_{j\sigma}
	]
	\nonumber \\
	&+&
	\sum_{\sigma'}
	{\mi}
	t'^{}_2
	(\hat{\boldsymbol{\sigma}} \!
	\times \hat{\br}^{}_{ij})_{\sigma\sigma'}^z
	[
	a^{\dagger}_{i\sigma}
	a^{}_{j\sigma'}
	-
	b^{\dagger}_{i\sigma}
	b^{}_{j\sigma'}
	]
	\Big\rbrace\, , \nonumber 
\end{eqnarray}
where
$\boldsymbol{\hat{\sigma}}=({\sigma}^{x},{\sigma}^{y},{\sigma}^{z})$
denotes the $2\times 2$ Pauli matrices in the pseudospin basis, $\mu$
is the chemical potential, and $a^{\dag}_{i\sigma}$
($b^{\dag}_{i\sigma}$) creates an electron at site $i$ in sublattice A
(B) with pseudospin $\sigma$. The parameter $t_2$ denotes the second
neighbor spin-independent (intra-sublattice) hopping integral and
$t^{}_1=t^{}_1(\theta)=\frac{2t_0}{3} \cos\theta (2 \cos^4\theta -1)$
and $t'^{}_1=t'^{}_1(\theta)=\frac{2t_0}{3} \sin\theta (2 \sin^4\theta
-1)$ are spin-independent and spin-dependent (inter-sublattice)
nearest-neighbor hopping amplitudes,
respectively~\cite{Jaejun_PRB_2009}. Here, the angle $\theta$
describes the rotation of the IrO$_6$ octahedra along the c-axis. The
last term in the non-interacting Hamiltonian stems from the second
neighbor spin-dependent (intra-sublattice) hopping with amplitude
$t'^{}_2$, and is a staggered ASOC that violates parity. The
combination of the IrO$_6$ rotation with the stacking structure of the
2D layers along the c-axis breaks the mirror symmetry with regard to
the ab-plane and results in the spin-dependent intra-sublattice
term~\cite{Yanase_PRL_2017}. Moreover, the crinkling of the lattice by
displacing the A (B) sublattice in the $\hat{z}~(-\hat{z})$ direction
allows for a second-neighbor (intra-sublattice) spin-dependent
hopping~\cite{Kane_PRL_2015}. To obtain the previously reported
electronic band
structure~\cite{Jaejun_PRB_2009,delaToore_PRL_2015,Ferrero_PRB_2018,Kee_PRB_2019},
we set the hopping parameters $t^{}_2=t^{2}_1$ and $t'_{2}=t'^{2}_1$,
respectively.  Furthermore, all of the physical parameters are scaled
in units of $t^{}_0=0.35$ eV to obtain a band structure in good
agreement with LDA+SOC
calculations~\cite{Jaejun_PRB_2009,delaToore_PRL_2015,Ferrero_PRB_2018}.

The repulsive interactions in Eq.~(\ref{ini_Ham}) are taken as a
combination of both on-site and extended Hubbard terms
%
\begin{equation}
{\cal H}^{}_{\rm Int}=
\frac{U}{2} 
\sum_{i,\sigma}
n^{}_{i\sigma}n^{}_{i\bar{\sigma}}
+
\frac{V}{2} 
\sum_{ \langle ij \rangle, \sigma\sigma'}
n^{}_{i\sigma}
n^{}_{j{\sigma'}},
\label{Eq:Hubbard_repulsion}
\end{equation}
%
where $U$ and $V$ are the strengths of the on-site and the
first-neighbor Hubbard interaction, respectively, 
 $n^{}_{i\sigma}$
is the pseudospin dependent electron occupation number operator, 
and we set $\bar{\sigma}=-{\sigma}$.

%
\begin{figure}[t]
	\begin{center}
		\hspace{-0.1cm} 
		\vspace{-0.35cm}
		\includegraphics[width=1.00 \linewidth]{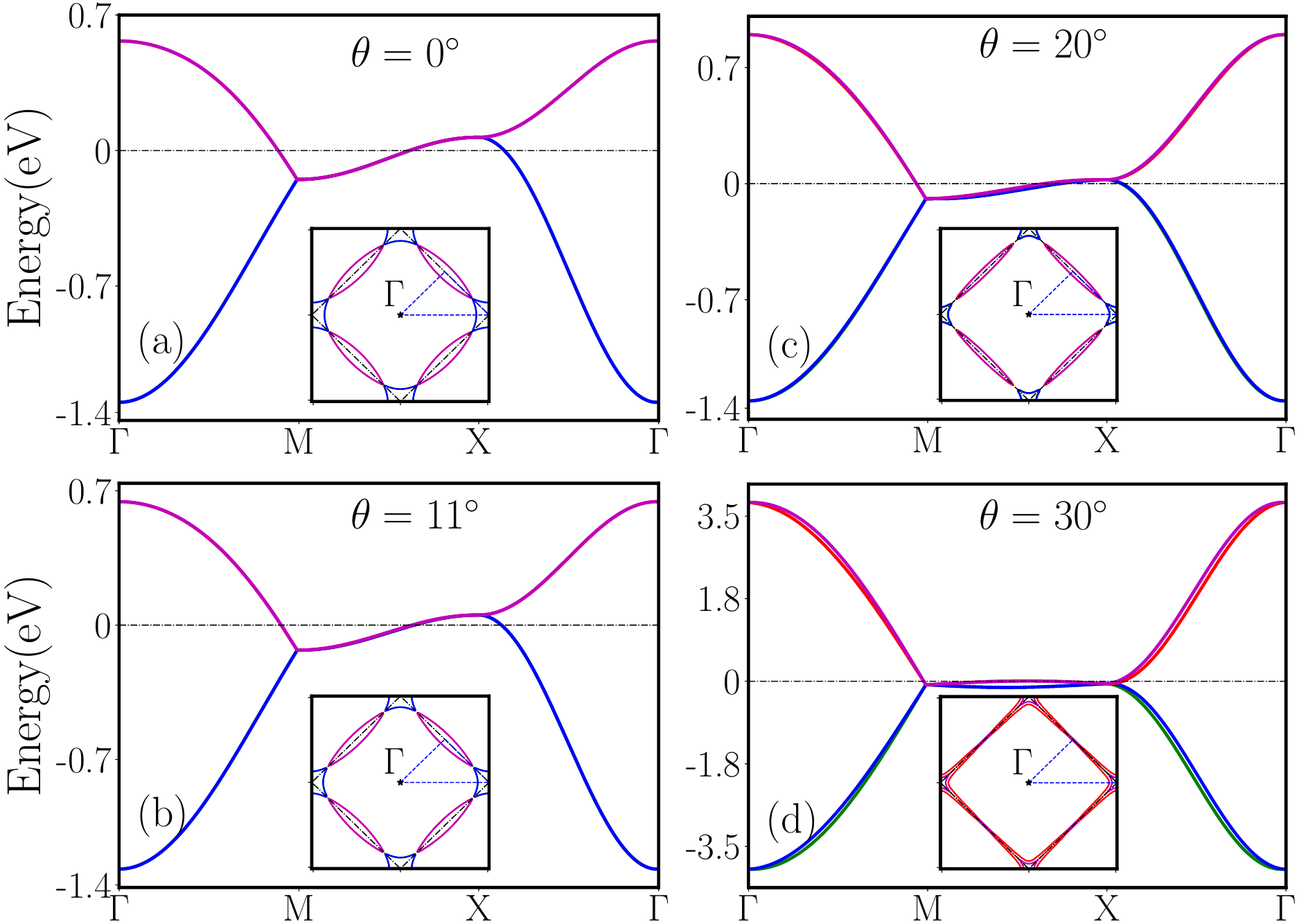} 
	\end{center}
	\caption{Evolution of the non-interacting band structure (main plots) 
		and the Fermi surface (insets) of the $J_{\rm eff}=1/2$ model
		\eqref{Eq:real_Ham} with respect to changes in the
		canting angle for the electron-doped case with $\langle n \rangle=2.4$.  
		The first BZ with the A and B sublattices is depicted as dashed lines.  
		Increasing the distortion of the Ir-O-Ir bonds will reduce the second neighbor hopping and ultimately restore the particle-hole symmetry of the band structure.}
	\label{Fig:BS_FS}
\end{figure}
%

Next, we perform a Fourier transformation of the fermionic
operators $a^{\dagger}_{\bk\sigma}=
\frac{1}{N}\sum_{i}
e^{{{\mi}} \bk\cdot \bR_i}
a^{\dagger}_{i\sigma} $,
and 
$
b^{\dagger}_{\bk\sigma}=
\frac{1}{N}\sum_{i}
e^{{{\mi}} \bk\cdot \bR_i} b^{\dagger}_{i\sigma} $ for the sublattices
A and B, respectively, with the number of points in momentum space
denoted by $N$. This leads to the Hamiltonian in reciprocal space,
which can be written as
%
\begin{align}
\begin{aligned}
{\cal H}^{}_0=
\sum_{\bk}
\Psi^{\dagger}_{\bk}
\begin{bmatrix}
\varepsilon^{}_{2\bk} 
+
\bg^{}_{2\bk}\!\!\cdot\!\boldsymbol{\sigma}
&\;
\varepsilon^{}_{1\bk} 
+
{\mi}\bg^{}_{1\bk}\!\!\cdot\!\boldsymbol{\sigma}
\\
\varepsilon^{}_{1\bk} 
-
{\mi}\bg^{}_{1\bk}\!\!\cdot\!\boldsymbol{\sigma}
&\;
\varepsilon^{}_{2\bk} 
-
\bg^{}_{2\bk}\!\!\cdot\!\boldsymbol{\sigma}
\end{bmatrix}
\Psi^{}_{\bk},
\end{aligned}    
\label{Eq:k_Ham}
\end{align}
%
where $\Psi^{\dagger}_{\bk}=(a^{\dagger}_{\bk\uparrow},a^{\dagger}_{\bk\downarrow},b^{\dagger}_{\bk\uparrow},b^{\dagger}_{\bk\downarrow})$, and
%
\bea
\bl
\varepsilon^{}_{1\bk}
&=
-4t^{}_1 
\Big[
\cos k_x+\cos k_y
\Big],
\label{Eq:1st_independent}
\\
\varepsilon^{}_{2\bk}
&=
-4t^{}_2 \cos k_x \cos k_y-\mu,
\label{Eq:2nd_independent}
\el
\eea
%
are the dispersions originating from the nearest-neighbor
(inter-sublattice) and next-nearest neighbor (intra-sublattice)
spin-independent hopping, respectively. Furthermore,
\begin{equation}
\bg^{}_{1\bk}=-4t'^{}_1 
\Big[
\cos k_x+\cos k_y
\Big]
\hat{z},
\label{Eq:1st_dependent}
\end{equation}
%
corresponds to the nearest-neighbor (inter-sublattice) spin-dependent
hopping. This term is a consequence of a deviation of the Ir-O-Ir bond
angle from $180^{\circ}$ which generates a quasi-SOC described by an
even vector $\bg^{}_{1\bk}$. Moreover,
%
\begin{equation}
\bg^{}_{2\bk}
\!
=
-4t'^{}_2 
\Big[
\sin k_x \cos k_y \hat{x}-\sin k_y \cos k_x \hat{y}
\Big],
\label{Eq:ASOC_g_vector}
\end{equation}
%
describes the ASOC $\bg$-vector.
Eqs.~(\ref{Eq:1st_dependent})~and~(\ref{Eq:ASOC_g_vector}) show that
only the ASOC results in a violation of parity. The energy dispersion
of the non-interacting normal system is therefore given by
%
\begin{align}
E^{}_{\bk,s}=
\!
-\mu+\varepsilon^{}_{2\bk}
+
\xi
\sqrt{
	\varepsilon^{2}_{1\bk}
	+
	(
	\bg^{}_{1\bk}
	+
	\xi'
	\:
	|\bg^{}_{2\bk}|
	)^2_{}
},
\label{Eq:Normal_Energy}
\end{align}
%
%
where $s=\lbrace 1,2,3,4 \rbrace$ denotes the band number
corresponding to $(\xi,\xi') = \lbrace(-1,-1), (-1,+1), (+1,-1),
(+1,+1)\rbrace$, respectively. The band filling is defined as the
number of electrons per unit cell and expressed as $\langle n
\rangle=2+2\rho$, in which $\rho$ corresponds to the doping level,
e.g. $\langle n \rangle=2$ for half-filling. It is worth mentioning
that at every specific level of doping, the value of the chemical
potential can be calculated from
%
%
\begin{equation}
\rho=\frac{1}{N}
\sum_{\bk,s}f(E_{\bk,s})-2,
\label{Eq:mu}
\end{equation}
%
 where $f(\ldots)$ 
 the Fermi-Dirac distribution function at temperature $T$. We are
 interested here in the limit $T\to 0$.
%
\begin{figure}[b]
	\includegraphics[width=1.0\columnwidth]{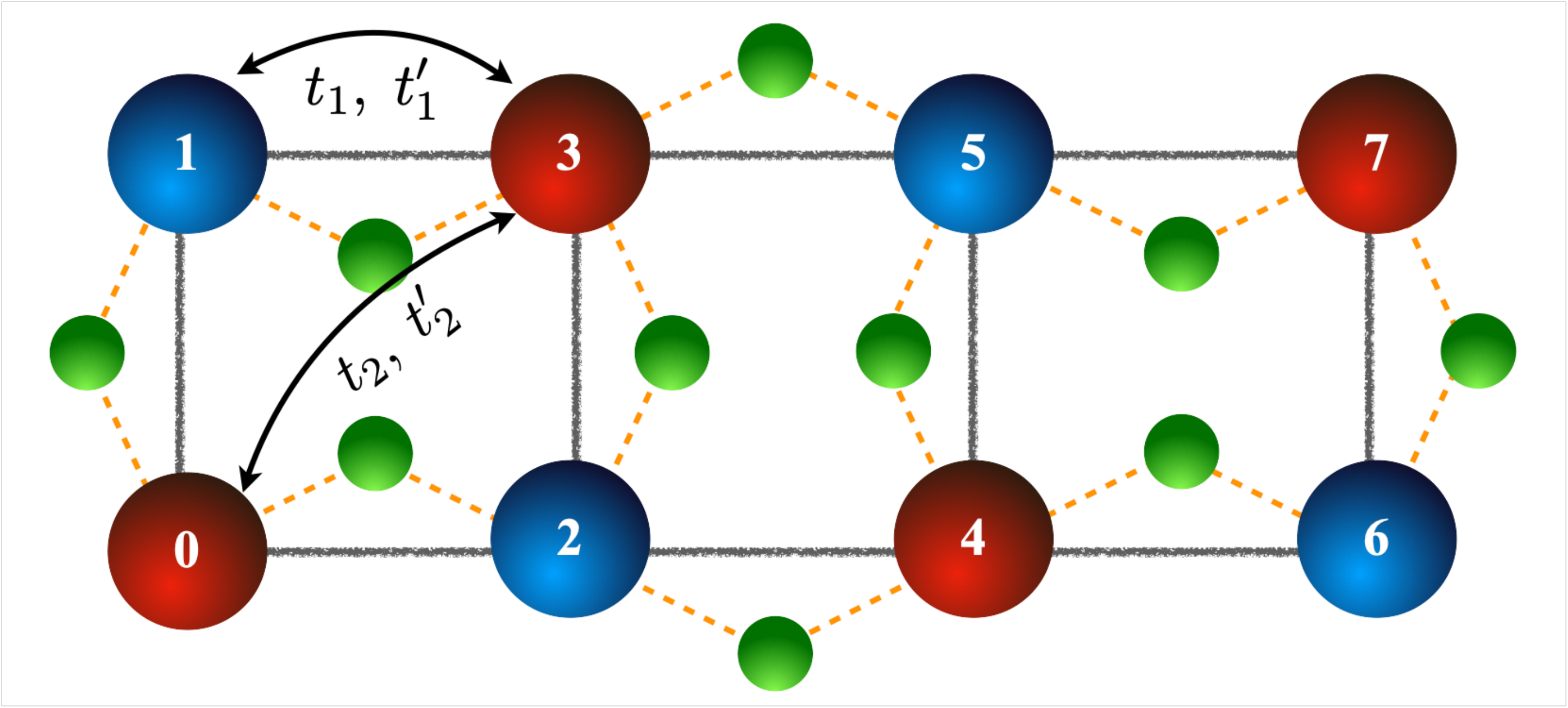}
	\caption{The basic building block of our model is a two-leg
		ladder with intra-sublattice hopping amplitudes $t_1$ and
		$t'_1$ and inter-sublattice hopping amplitudes $t_2$ and
		$t'_2$. We consider an onsite Hubbard interaction $U$ and a
		nearest-neighbor interaction $V$.}  
	\label{Fig_Ladder}
\end{figure}
%
\begin{figure*}[t]
	\includegraphics[width=1.0\textwidth]{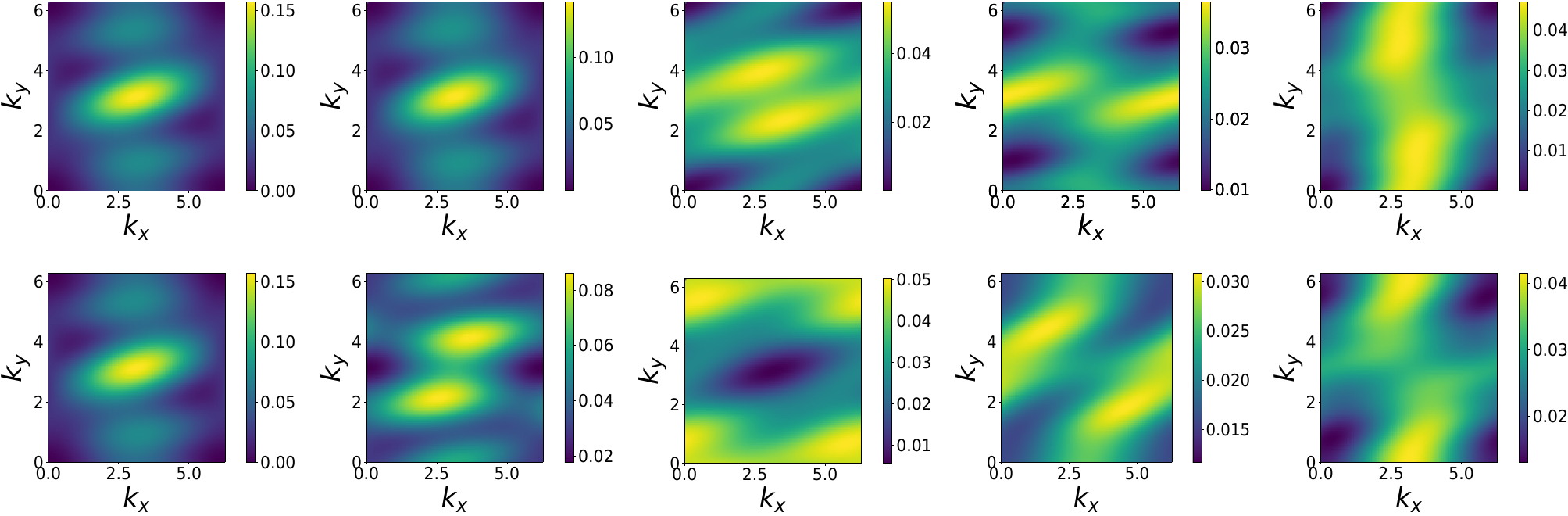}
	\caption{Magnetic structure as obtained from exact diagonalizations of small clusters. In all cases $U=4$ and $V=0.4$. Upper row, left to right: static longitudinal spin structure factors for a $2\times 4$ cluster with $\langle n\rangle =2$ and $\theta = 0^{\circ},20^{\circ},30^{\circ}$ followed by the results for a $2\times 5$ cluster with $\theta=20^{\circ}$ and $\langle n\rangle =2.8$ and finally $\langle n\rangle =1.2$. The lower row shows the corresponding transverse structure factors.}
	\label{Fig_ED}
\end{figure*}
%

Let us now discuss the evolution of the non-interacting band structure
and the Fermi surface with respect to the rotation of the IrO$_6$
octahedra around the $z$ axis at a specific filling of $\langle n
\rangle = 2.4$. Fig.~\ref{Fig:BS_FS}(a) shows the case of canting
angle $\theta=0^\circ$. In this case, the band structure and Fermi
surface is two-fold degenerate due to the absence of both first- and
second-neighbor spin-dependent hopping. Figs.~\ref{Fig:BS_FS}(b-d)
show the effect of an increase in the canting angle on the band
structure and Fermi surface. In particular, the splitting between the
bands increases with increasing angle $\theta$. Moreover, an
increasing canting angle also reduces the amplitude of the
next-nearest neighbor spin-independent hopping. This leads to almost
particle-hole symmetric bands at the largest canting angle
$\theta=30^\circ$ shown in Fig.~\ref{Fig:BS_FS}(d). At the same time,
semi-flat bands and line nodes are formed along the Brillouin zone
(BZ) boundary from the M to the X high-symmetry points. The
corresponding evolution of the Fermi surface is shown in the insets of
Fig.~\ref{Fig:BS_FS}.

\section{Exact results for small clusters}
\label{Sec_ED}
In order to gain some insights into the interplay of the various terms
in Eq.~\eqref{Eq:real_Ham} and the Coulomb interactions
\eqref{Eq:Hubbard_repulsion}, we start by considering small clusters
using exact diagonalizations. As the central building block, we
consider a two-leg ladder of alternating A and B sublattice sites
which constitutes `two rows' in the lattice shown in
Fig.~\ref{Fig:Unit_Cell}. This two-leg ladder---including the various
hopping terms---is also shown more schematically in
Fig.~\ref{Fig_Ladder}. Because spin is not a good quantum number due
to the $t'_2$ term, we only have particle number conservation to limit
the size of the Hilbert space thus restricting our exact
diagonalizations to relatively small cluster sizes. Nevertheless, as
we show below, these cluster sizes are sufficient to gain some
valuable insights.

We concentrate on intermediate interaction strengths $U=4$ and $V=0.4$
and checked that the results are qualitatively very similar for $U=2$
to $U=8$ with $V=U/10$. Furthermore, for $V\ll U$, as is expected for
Sr$_2$IrO$_4$, there will be no charge order. We thus focus entirely
on the magnetic order as a function of canting angle $\theta$ and
doping level $\langle n\rangle$. We consider a two-leg ladder with
open boundary conditions and calculate the correlation functions
$\langle S^z_0 S^z_r\rangle$ and $\langle S^+_0 S^-_r\rangle$ where
the sites are numerated as shown in Fig.~\ref{Fig_Ladder}. A Fourier
transform then leads to the static spin structure factors 
{
$S^{zz}(\bk)$
}
and $S^{+-}(\bk)$ which we use as our main observables to discuss the
possible magnetic orderings. While dynamical structure factors can be
calculated as well, we find that the accessible cluster sizes are too
small to learn much about the dispersion of the magnetic excitations.

In Fig.~\ref{Fig_ED}, results for the static spin structure factors are
shown for different canting angles and filling fractions. We start by
considering the half-filled case, $\langle n\rangle =2$. For
$\theta=0^{\circ}$, the model has $SU(2)$ spin rotational symmetry and
the longitudinal and transverse spin correlations are identical. For
this case we find strong antiferromagnetic correlations, see the
leftmost column in Fig.~\ref{Fig_ED}. There is a peak in the structure
factor centered at $(\pi,\pi)$. Increasing the canting angle in the
half-filled case, the magnetic structure remains largely unchanged up
to $\theta\lesssim 20^{\circ}$ although transverse and longitudinal
spin correlators are, of course, no longer exactly equal. For the case
$\theta=20^{\circ}$, shown in the second column of Fig.~\ref{Fig_ED},
$S^{+-}(\bk)$ shows a splitting of the antiferromagnetic peak into two
incommensurate peaks. For $\theta=30^{\circ}$, shown in the third
column of Fig.~\ref{Fig_ED}, most of the spectral weight in
$S^{+-}(\bk)$ has moved to $\bk=(0,0)$ (and points equivalent by a
reciprocal lattice vector), i.e., the in-plane magnetic correlations
are now ferromagnetic. In the longitudinal direction, on the other
hand, the correlations remain weakly antiferromagnetic. The peak at
$(\pi,\pi)$, however, is now slightly split indicating incommensurate
correlations.

This transition at half-filling from fully antiferromagnetic
correlations to ferromagnetic in-plane and antiferromagnetic
out-of-plane correlations with increasing canting angle can be
understood as follows: The $t'_2$ term in the Hamiltonian
\eqref{Eq:real_Ham} describes spin-flip hopping and therefore
kinematically favors a ferromagnetic alignment of spins. At the same
time, the other hopping terms still prefer an antiferromagnetic
alignment. As the importance of the $t'_2$ term increases with
increasing canting angle, the system compromises by developing
in-plane ferromagnetic correlations whereas the out-of-plane
correlations, while weakened, remain antiferromagnetic.

Next, we investigate the changes to the magnetic structure when doping
the system. Here we keep $\theta=20^{\circ}$ fixed and consider both
the electron and the hole doped case. For the electron doped case,
$\langle n \rangle = 2.8$, shown in the fourth column of
Fig.~\ref{Fig_ED}, we find that the longitudinal correlation function
now has a peak at $(0,\pi)$, i.e., the correlations are ferromagnetic
along the legs but antiferromagnetic between the two legs. The
transverse correlations appear to be largely incommensurate. In
contrast, the hole doped case $\langle n\rangle = 1.2$ shown in the
last column of Fig.~\ref{Fig_ED} has dominant $(\pi,0)$ correlations:
antiferromagnetic along the legs and ferromagnetic between the two
legs, both in- and out-of-plane. In both cases the peaks are less
well-defined than in the undoped case and the magnetic order is much
weaker.

Clearly, all these results are affected by the small cluster sizes and
related boundary effects. Nevertheless, there are two main conclusions
we can draw in the doped case: (1) the results are strongly
particle-hole asymmetric, and (2) doping does weaken the magnetic
correlations and also tends to push them to become more incommensurate
with the lattice.

In conclusion, we have gained important qualitative insights into the
model Hamiltonian \eqref{Eq:real_Ham} for intermediate interaction
strengths. At half-filling, the antiferromagnetic order is weakened
with increasing canting angle, ultimately leading to a transition to
ferromagnetic in-plane correlations for $\theta \gtrsim
20^{\circ}$. Doping the system further weakens the magnetic order
while making the correlations also more incommensurate. As expected,
the cases of hole and electron doping are not equivalent. Based on the
exact diagonalizations of such small clusters, we cannot make any
statements about how long-range these magnetic structures are and what
happens if such a basic building block is coupled to its
surroundings. To address these questions, we calculate the magnetic
response for the full two-dimensional lattice using the random-phase
approximation next.


\section{Random-phase approximation}
\label{Sec_RPA}
In the single-particle response, the effects of correlations weaken
rapidly with doping, such that one may expect the random phase
approximation to provide an adequate description of the two-particle
response. However, this is not necessarily true as has been shown for
the case of the cuprates in Ref.~\cite{Devereaux_PRB_2015}. Since
we are here interested in a much more moderately correlated regime as
compared to the cuprates we might, however, nevertheless expect that
RPA provides a good starting point to analyze the basic physical
properties of this model.
\subsection{Response functions}
In the framework of linear response theory and using the Kubo formula,
the physical components of the bare susceptibilities are given
by~\cite{Cobo_PRB_2016}
%
\begin{align}
\begin{aligned}
&
\chi_{0}^{uv}(\bq,{{\mi}}\omega_n)=
\\&
-\frac{T}{4N}
\sum_{\bk,{{\mi}}\nu_m}
{\rm Tr}
\Big[
\check{\boldsymbol \sigma}^{u}
\check{G}^{}_{0}
(\bk,{\mi}\nu_m)
\check{\boldsymbol \sigma}^{v}
\check{G}^{}_{0}
(\bk+\bq,{{\mi}}\nu_m+{{\mi}}\omega_n)
\Big],
\end{aligned}
\label{Eq:Kubo_bare}
\end{align}
%
in which $u,v=0$, and $u,v=\lbrace x,y,z \rbrace$ are denoting the
charge and spin components of the bare susceptibility. Moreover, the
trace is performed over sublattice and spin spaces and
%
\begin{align}
\check{\boldsymbol \sigma}^{u}
=
\begin{cases}
\mathbbm{1} 
& u=0
\\
\tau^z \otimes \sigma^u  & u\neq 0
\end{cases},
\label{Eq:Sigma}
\end{align}
%
 where $\mathbbm{1}$ is a $4\times 4$ unit matrix and
 $\boldsymbol{\hat{\tau}}=({\tau}^{x},{\tau}^{y},{\tau}^{z})$ are
 Pauli matrices in sublattice space. Besides, the unperturbed electron
 Green's function in the same basis is defined as
%
\begin{align}
\begin{aligned}
\check{G}^{}_{0}
(\bk,{\mi}\nu_m)
=
\Big[
{\mi}\nu_m \mathbbm{1}-{\cal H}_0(\bk)
\Big]^{-1}
\end{aligned}
\label{Eq:Free_Green}
\end{align}
where $\nu_m$ are Matsubara frequencies. One can write the unperturbed
Green's function matrix as
%
\begin{align}
\check{G}^{}_{0}
(\bk,{{\mi}}\nu_m)
=
\begin{bmatrix}
\hat{G}^{\rm AA}_{0}(\bk,{{\mi}}\nu_m) & \hat{G}^{\rm AB}_{0}(\bk,{{\mi}}\nu_m)
\\
\hat{G}^{\rm BA}_{0}(\bk,{{\mi}}\nu_m) & \hat{G}^{\rm BB}_{0}(\bk,{{\mi}}\nu_m)
\end{bmatrix}.
\label{Eq:G}
\end{align}
%
Then, the bare susceptibility within the sublattice-spin basis can be
rewritten as
%
\begin{align}
\begin{aligned}
\chi_{0}^{uv}(\bq,{{\mi}}\omega_n)
&
=
-\frac{T}{4N}
\sum_{\bk,{{\mi}}\nu_m}
\sum_{pp'p''}
\\&
\hspace{-0.4cm}
{\rm Tr}^{}_{\sigma}
\Big[
\hat{\sigma}^u
\hat{G}^{pp'}_{0}(\bk,{{\mi}}\nu_m)
\hat{\sigma}^v
\hat{G}^{p'\!p''}_{0}(\bk+\bq,{{\mi}}\nu_m+{{\mi}}\omega_n)
\Big].
\end{aligned}
\label{Eq:New_bare}
\end{align}
%
The transformation of the free electron Green's function from the
sublattice-spin into the band pseudospin basis is achieved by
%
\begin{align}
\begin{aligned}
G^{pp'}_{0,\sigma\sigma'}(\bk,{{\mi}}\nu_m)
=\sum_{s}
\Lambda^{s}_{p\sigma}(\bk)
\Lambda^{*s}_{p'\!\sigma'}(\bk)
G^{s}_{0}(\bk,{{\mi}}\nu_m),
\end{aligned}
\label{Eq:Green_Transformation}
\end{align}
%
where the number of bands in Eq.~(\ref{Eq:Normal_Energy}) is
represented by $s = 1,2,3,4$, and $\Lambda^{s}_{p\sigma}(\bk) = \bra
\bk,p\sigma|\bk,s\ket$ denotes the matrix elements to connect the $s$-th
band to sublattice $p$~(=A or B) and pseudospin $\sigma$.
Hence, the spatial components of the bare susceptibility are given by
%
\begin{align}
\begin{aligned}
&
\chi_{0}^{uv}(\bq,{{\mi}}\omega_n)
=
\\
&
-\frac{T}{4N}
\!\!
\sum_{\bk,ss',{{\mi}}\nu_m}
\!\!\!\!
\zeta^{ss'}_{uv}(\bk,\bq)
G^{s}_{0}(\bk,{{\mi}}\nu_m)
G^{s'}_{0}(\bk+\bq,{{\mi}}\nu_m+{{\mi}}\omega_n),
\end{aligned}
\label{Eq:Bare_Susc_1}
\end{align}
%
with
%
\bea\nonumber
\begin{aligned}
&
\zeta^{ss'}_{uv}\!(\bk,\bq)
\!
=
\\
&\;\;\;
\Lambda^{*s'}_{p\sigma}(\bk+\bq)
\sigma^{u}_{\sigma \sigma'}
\Lambda^{s}_{p'\sigma'}(\bk)
\;
\Lambda^{s'}_{p'\!\delta}(\bk+\bq)
\sigma^{v}_{\delta' \delta}
\Lambda^{*s}_{p''\!\delta'}(\bk)
,
\end{aligned}
\label{Eq:zets}
\eea
%
where a summation is performed over the repeated spin-indices. If we
now sum over the fermionic Matsubara frequency ${{\mi} \nu_m}$ and
do an analytical continuation ${{\mi}}\omega_n\rightarrow
\omega+{{\mi}}0^{+}_{}$ then we obtain the well-known Lindhard function
for the retarded bare susceptibility
%
%
\begin{align}
\begin{aligned}
\chi_{0}^{uv}(\bq,\omega)=
\frac{1}{4N}
\sum_{\bk,ss'}
\!
\zeta^{ss'}_{uv}(\bk,\bq)
\frac{
f(E^{s'}_{\bk+\bq})-f(E^{s}_{\bk})
}
{
E^{s}_{\bk}-E^{s'}_{\bk+\bq}+\omega+{{\mi}}0^{+}_{}
}.
\end{aligned}
\label{Eq:Bare_Susc_Lindhard}
\end{align}
%
%
Within RPA, the matrix of susceptibilities is then given by
%
\begin{equation}
\hat{\chi}_{\rm RPA}^{}(\bq,\omega)=
\frac
{1}
{
1-\hat{U}(\bq)
\hat{\chi}^{}_{0}(\bq,\omega)
}
\hat{\chi}^{}_{0}(\bq,\omega),
\label{Eq:RPA_Kappa}
\end{equation}
%
where $\hat{U}(\bq)$ denotes the bare interaction matrix with
%
\begin{align}
\hat{U}(\bq)=
\sum_{uv}
\delta^{}_{u,v}
\Big[
\delta^{}_{u,0}
V(\bq)
-
(-1)^{\delta^{}_{u,0}}\frac{U}{8}
\Big].
\label{Eq:Int_Matrix}
\end{align}
%
In this equation, $\delta^{}_{u,v}$ is the Kronecker delta and
$V(\bq)=2 V (\cos q_x+\cos q_y)$. In the spin (charge) channel at a
specific value of $U=U_c$ ($V=V_c$), the determinant of the
denominator of Eq.~(\ref{Eq:RPA_Kappa}) vanishes,
$|1-\hat{U}(\bq)\hat{\chi}_{0}^{}(\bq,\omega)|=0$, generating an
instability towards an ordered spin-density wave (charge-density wave)
state~\cite{Ghadimi_PRB_2019}.

In addition to the zero-frequency spin and charge susceptibilities, we
are also calculating the charge-charge and spin-spin two-point
correlation functions which are obtained by the Fourier transformation
of the charge and spin susceptibility, respectively. The
density-density correlation function is expressed as
%
\begin{equation}
\langle n({\bf 0}) n(\br) \rangle =
\frac{1}{4N}
\sum_{\bq}
e^{{\mi}\bq \cdot \br} 
\,
\chi_{\rm RPA}^{00}(\bq,\omega=0).
\label{Eq:n0_n0}
\end{equation}
%
Furthermore, the in-plane and out-of-plane components of spin-spin
correlation function are given by
\bea
\bl
\langle S^{+}({\bf 0}) S^{-}(\br) \rangle
&=
\frac{1}{4N}
\sum_{\bq}
e^{{\mi}\bq \cdot \br}
\,
 \chi_{\rm RPA}^{+-}(\bq,\omega=0),
\label{Eq:Sp_Sm1}
\\
\langle S^{z}({\bf 0}) S^{z}(\br) \rangle
&=
\frac{1}{4N}
\sum_{\bq}
e^{{\mi}\bq \cdot \br}
\,
 \chi_{\rm RPA}^{zz}(\bq,\omega=0),
\hspace{1cm}
\label{Eq:Sz_Sz}
\el
\eea
%
where
\begin{eqnarray}
&&
\chi_{\rm RPA}^{+-}(\bq,\omega)
=
\\
&&
\chi_{\rm RPA}^{xx}(\bq,\omega)
\!+\!
\chi_{\rm RPA}^{yy}(\bq,\omega)
-{\mi}
\Big[
\chi_{\rm RPA}^{xy}(\bq,\omega)
\!-\!
\chi_{\rm RPA}^{yx}(\bq,\omega)
\Big]
\nonumber
\end{eqnarray}
describes the in-plane component of the spin susceptibility
corresponding to spin flipping processes.

We also consider the dynamical structure factor $S_{\rm
RPA}^{uv}(\bq,\omega)$. Technically, we follow earlier
studies~\cite{Jesko_IOP_2007,Alireza_PRB_2014} for calculating the
different branches of spin excitations within RPA and use the formula
%
\begin{align}
\begin{aligned}
S_{\rm RPA}^{uv}(\bq,\omega)=
-2\,
{\rm Im}
\Big[
\chi_{\rm RPA}^{uv}(\bq,\omega)
\Big].
\end{aligned}
\label{Eq:Magnon}
\end{align}
%

%

\subsection{Results}
\begin{figure}[t]
	\begin{center} \hspace{.01cm} \includegraphics[width=0.90
	\linewidth]{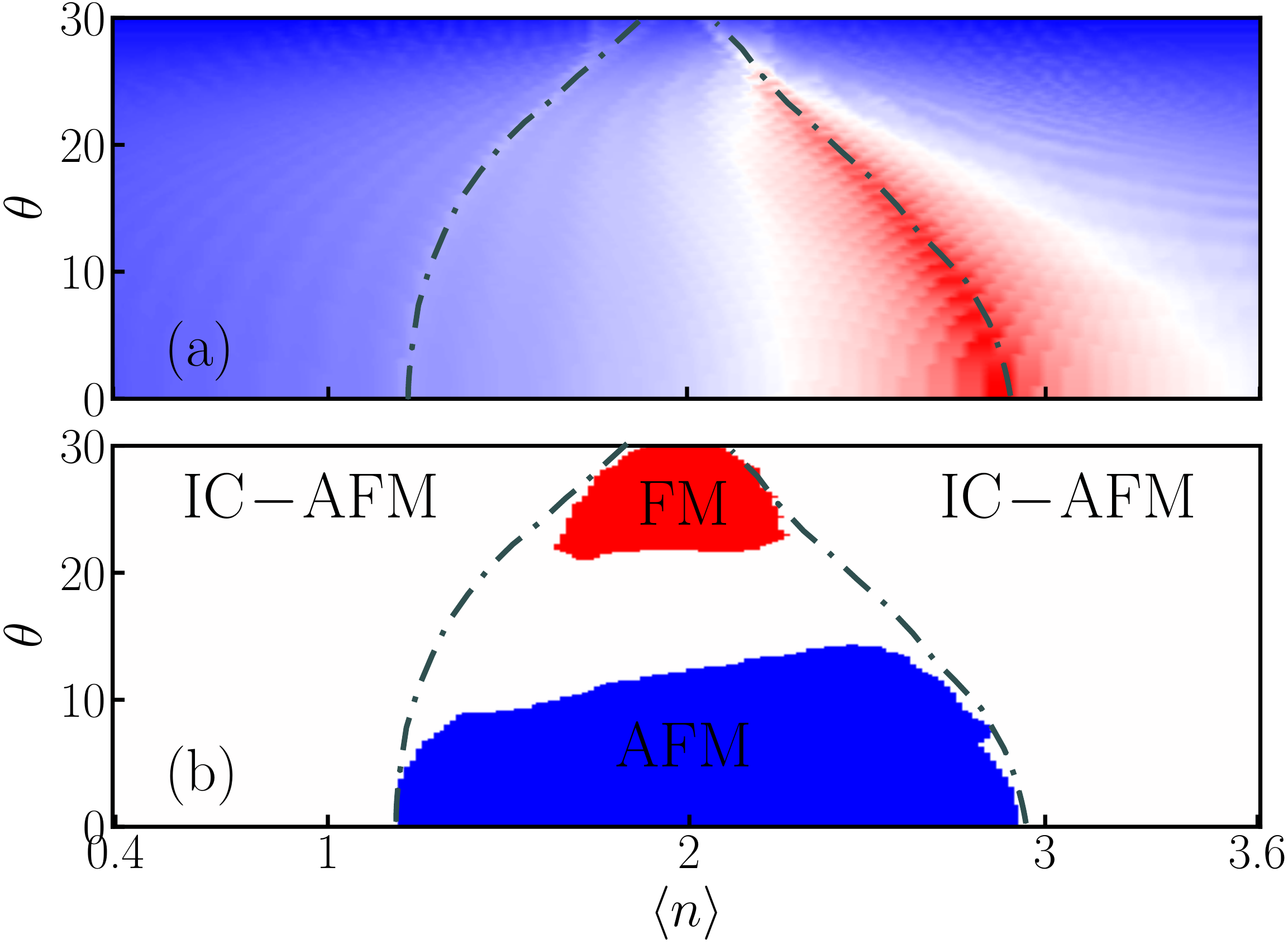} \vspace{-0.8 cm} %
	\end{center}  
\caption{(a) Density of states at the Fermi surface
	($\omega=0$) of the non-interacting model versus filling and
	canting angle.  The lines denote the position of the van-Hove
	singularities for specific values of filling, $\langle n
	\rangle^{}_{\rm vH}$, and distortions of the Ir-O-Ir
	bonds. (b) Magnetic phase diagram obtained by RPA as a
	function of filling and canting angle for $U=1$ and $V=0$. The
	lines denote again the position of the van-Hove singularities.}
\label{Fig:FS_DOS}
\end{figure}

We start by presenting and discussing our main result, the magnetic
phase diagram as a function of doping and canting angle, shown in
Fig.~\ref{Fig:FS_DOS}. We note first that in the density of states
(DOS) at the Fermi level there is an asymmetry in the position of the
van-Hove filling $\langle n \rangle^{}_{\rm vH}$ due to the non-zero
next-nearest neighbor hopping, see Fig.~\ref{Fig:FS_DOS}(a). By
increasing the canting angle, the van-Hove filling approaches half
filling, $\langle n \rangle=2$. The position of the van-Hove
singularities can help us to understand the transitions in the nature
of the magnetic fluctuations~\cite{Greco_PRB_2020}.  %
%
Fig.~\ref{Fig:FS_DOS}(b) shows the magnetic phase diagram of the
system for $U = 1$ obtained within RPA. Since the extended Hubbard
term is expected to be small and will thus only affect the charge
fluctuations, we set $V = 0$ for now and return to the effects of a
finite $V$ later. At low levels of canting angle, near
$\theta=0^{\circ}$, there are two distinguishable van-Hove fillings
near $\langle n
\rangle^{}_{\rm vH,1} \approx 1$ and $\langle n
\rangle^{}_{\rm vH,2} \approx 3$. The interval 
$\langle n \rangle^{}_{\rm vH,1} \lesssim \langle n \rangle \lesssim
\langle n \rangle^{}_{\rm vH,2}$ includes the area of commensurate
antiferromagnetic (AFM) fluctuations in the system. Outside of this
region, incommensurate AFM (IC-AFM) fluctuations dominate. Increasing
the canting angle reduces the region in doping with commensurate AFM
fluctuations. This can be explained by the van-Hove singularities
moving towards half filling. Finally, near half filling and for
canting angles $\theta\gtrsim 20^\circ$, ferromagnetic fluctuations are
established. Note that these fluctuations are not necessarily long
ranged as we will discuss in detail in the following.

%
\begin{figure}
  \centering 
\includegraphics[width=1.012 \linewidth, angle=0]{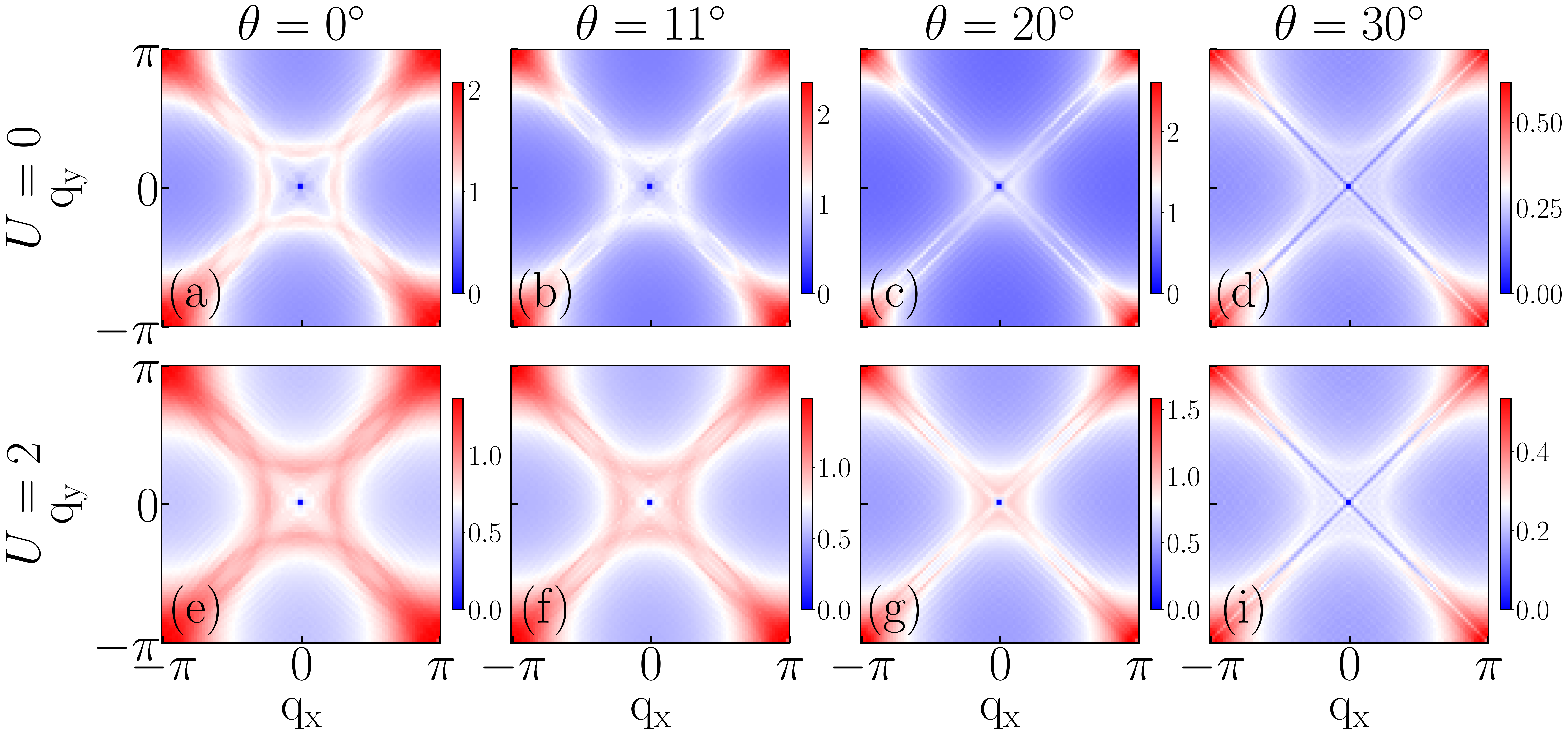} 
\caption{Momentum dependence of the zero-frequency
  charge susceptibility $\chi_{}^{00}(\bq,\omega=0)$ for $\langle n
  \rangle\!\!=\!\!2.4$ and different values of the canting angle. The
  top row shows the bare susceptibilities, and the bottom the
  corresponding RPA susceptibilities for $U=2$. We set $V=0$. The
  values of the RPA charge susceptibilities remain positive, i.e.~the
  system is far from a CDW instability.}
\label{Fig:Kappa_00_U_theta}
\end{figure}
%
In Fig.~\ref{Fig:Kappa_00_U_theta}, the evolution of the
zero-frequency charge susceptibility $\chi_{}^{00}(\bq,\omega=0)$ is
shown in the electron doped case as a function of canting angle. As
might be expected, we find that the on-site Hubbard interaction has
little effect; the data shown in the bottom row of
Fig.~\ref{Fig:Kappa_00_U_theta} are representative for $U\leq
3$. Charge fluctuations with $\bq\sim (\pi,\pi)$ are dominant but
remain always short-ranged---we are far from a CDW instability.

\begin{figure}
  \centering
\includegraphics[width=1.012 \linewidth, angle=0 ]{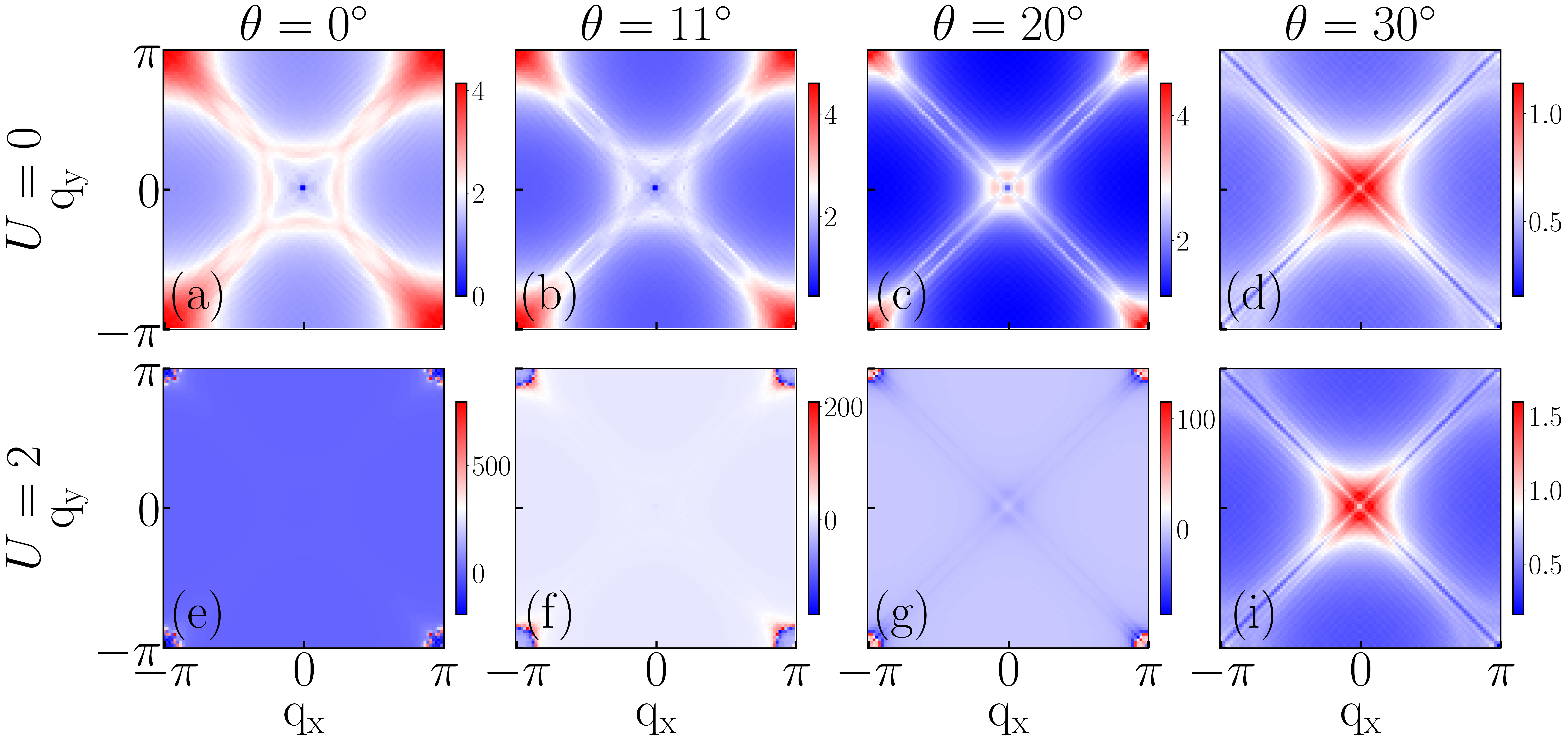} 
\caption{Momentum dependence of the in-plane component of the zero-frequency 
	spin susceptibility $\chi_{}^{+-}(\bq,\omega=0)$ for $\langle n
	\rangle\!\!=\!\!2.4$ and different values of $\theta$. The panels in the
	top row show the bare susceptibilities while the corresponding RPA
	results are presented in the bottom row. We have set $V=0$.}
\label{Fig:Kappa_pm_U_theta}
\end{figure}

The momentum structure of the zero-frequency in-plane spin
susceptibility as a function of $U$ and $\theta$ is presented in
Fig.~\ref{Fig:Kappa_pm_U_theta}. The bare susceptibilities
$\chi_{0}^{+-}(\bq,\omega=0)$, shown in
Figs.~\ref{Fig:Kappa_pm_U_theta}(a)-(d), display nearly commensurate
AFM fluctuations, $\bq\sim (\pi,\pi)$, for $\theta = 0^{\circ},
11^{\circ}, 20^{\circ}$ but nearly ferromagnetic fluctuations for
$\theta=30^{\circ}$. These fluctuations are, however, all short
ranged. For $U=2$, the RPA susceptibilities for
$\theta=0^{\circ},11^{\circ},20^{\circ}$ become negative, indicating
that long-range antiferromagnetic has been established. For
$\theta=30^\circ$ the RPA susceptibility is then again very similar to
the bare one: the ferromagnetic fluctuations remain short ranged.
%

\begin{figure}
  \centering
\includegraphics[width=1.012 \linewidth, angle=0 ]{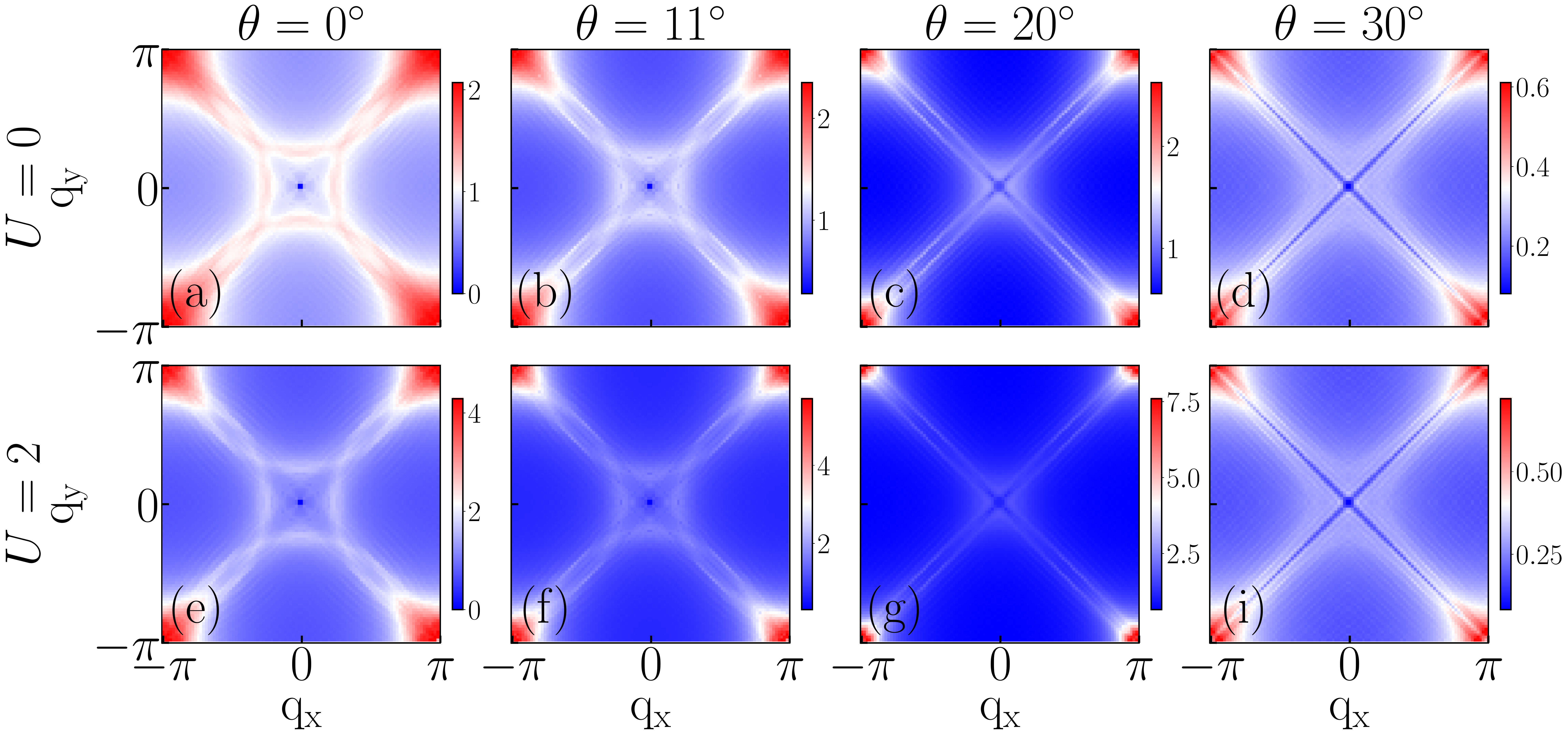} 
\caption{Same as Fig.~\ref{Fig:Kappa_pm_U_theta} but for the out-of-plane 
spin susceptibility $\chi_{}^{zz}(\bq,\omega=0)$.}
\label{Fig:Kappa_zz_U_theta}
\end{figure}

%
In Fig.~\ref{Fig:Kappa_zz_U_theta} we show the corresponding
out-of-plane spin susceptibilities. Here we find that for all canting
angles commensurate, short-ranged antiferromagnetic fluctuations are
present. The Hubbard interaction has, in this case, only a small
effect. The RPA susceptibilities are very similar to the bare ones.
%

\begin{figure}[t]
\begin{center}
\vspace{-0.16cm}
\includegraphics[width=.91 \linewidth]{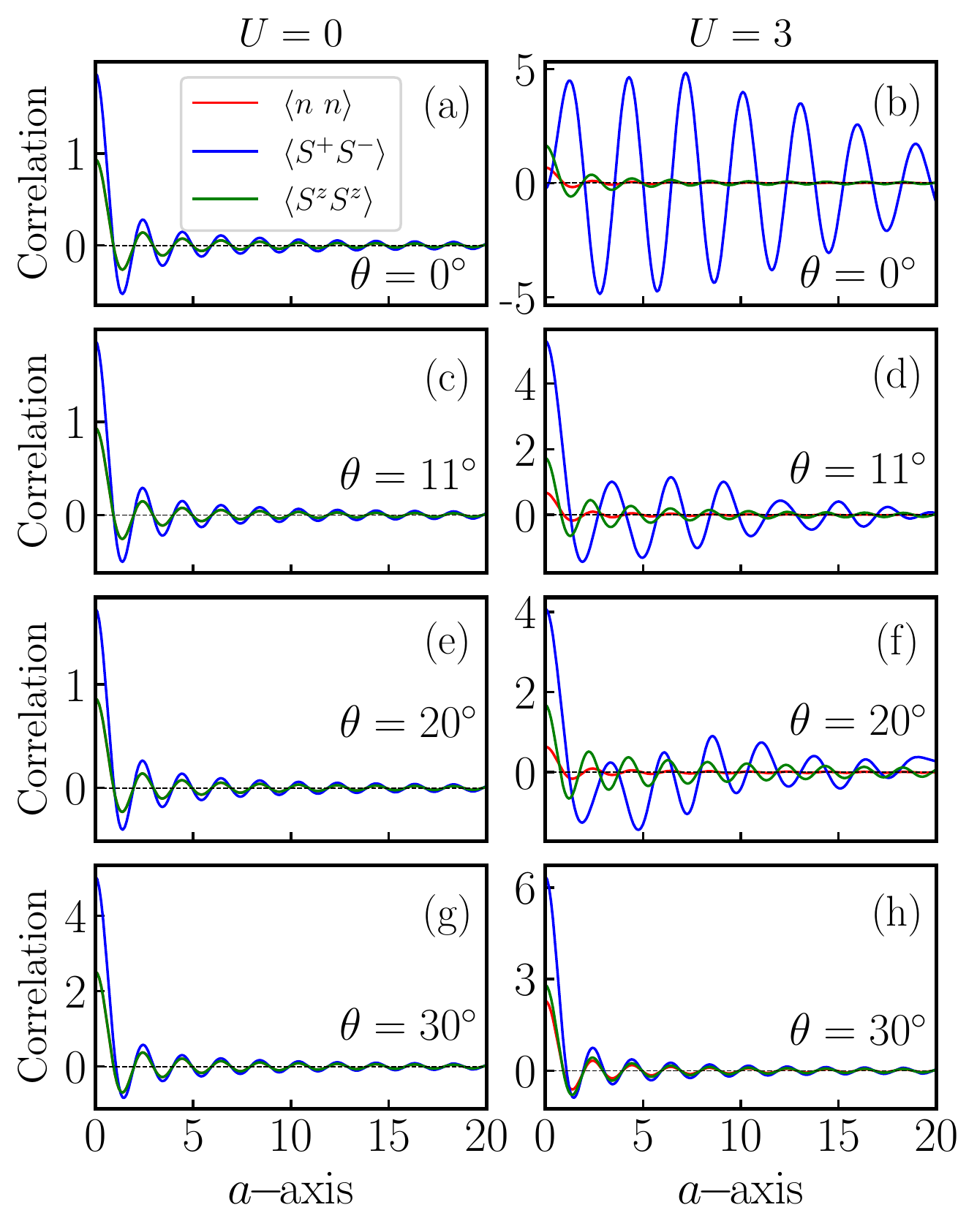}
\vspace{-.9cm}
\end{center}
\caption{Charge-charge (red), in-plane spin-spin (blue), 
and out-of-plane spin-spin (green) two-point correlation functions
along the crystallographic $a$-axis for the electron-doped case with
$\langle n \rangle = 2.4$. The left column shows the non-interacting
case and the right column the case with $U=3$ for different canting
angles $\theta$. We set $V = 0$.}
\label{Fig:Correlation_U_theta}
\end{figure}

%
To summarize these findings for the electron-doped case, we show in
Fig.~\ref{Fig:Correlation_U_theta} the zero-frequency two-point spin
and charge correlation functions along the crystallographic
$a$-axis. 
The
out-of-plane spin correlations $\langle S^z({\bf 0})S^z(\br)\rangle$
are always short-ranged and close to commensurate irrespective of the
Hubbard interaction or the canting angle. For the in-plane
correlations $\langle S^+({\bf 0})S^-(\br)\rangle$ we do see, on the
other hand, large changes as a function of the Hubbard interaction $U$
and the canting angle $\theta$. For $\theta=0^{\circ}, 11^{\circ},
20^{\circ}$ the Hubbard interaction establishes long-range
antiferromagnetic correlations in the system. These correlations are
fully commensurate for $\theta=0^{\circ}$ and become slightly
incommensurate for $\theta= 11^{\circ}, 20^{\circ}$. For all canting
angles with and without interactions the charge-charge correlations
remain always commensurate and short ranged. For pure on-site Hubbard
interactions there is no CDW instability.
%

\begin{figure}
  \centering
\includegraphics[width=1.012 \linewidth, angle=0 ]{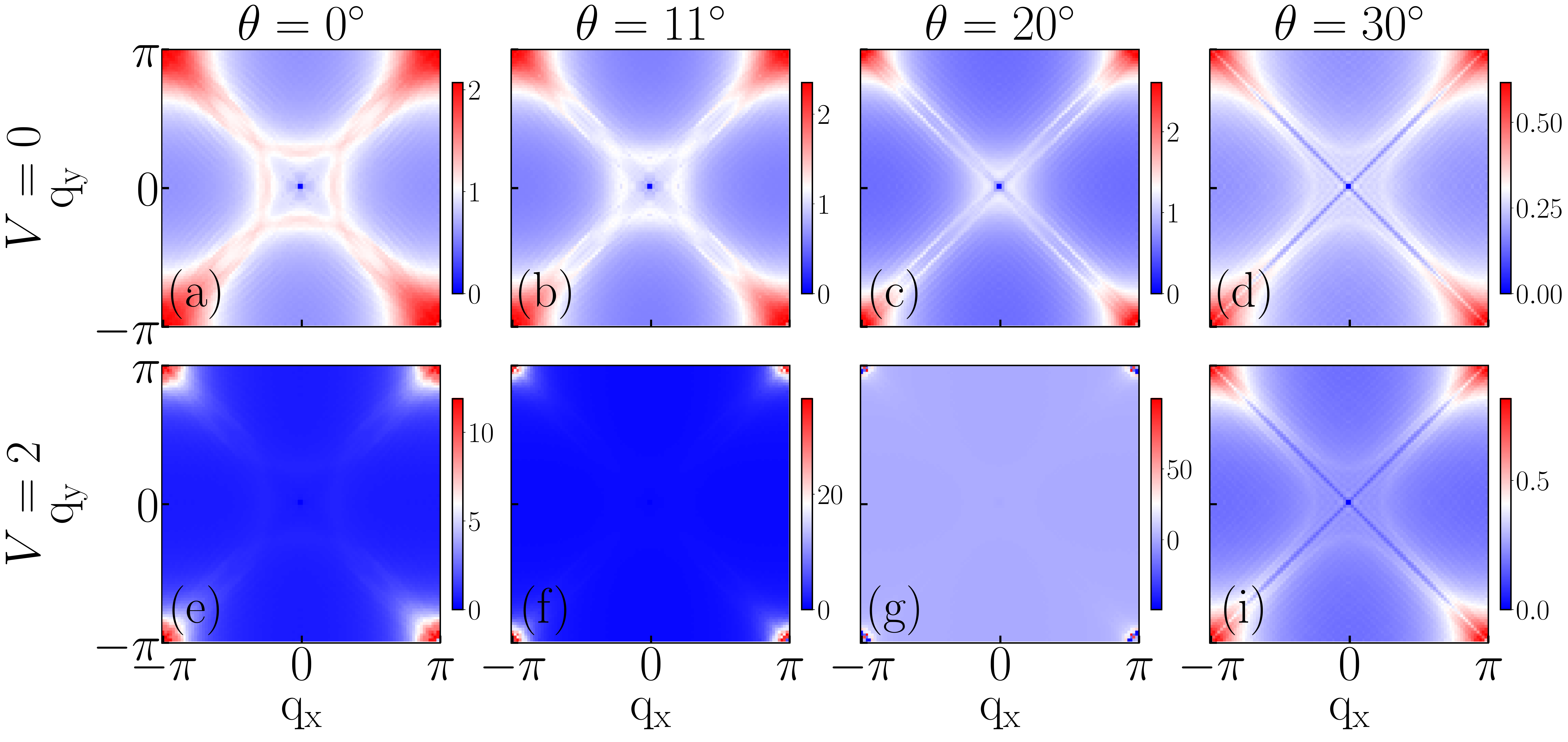} 
\caption{Charge susceptibility $\chi_{}^{00}(\bq,\omega=0)$ 
for the same parameters as shown in Fig.~\ref{Fig:Kappa_00_U_theta}
but now we set $U=0$ and compare the non-interacting case with the
case of weak nearest-neighbor Coulomb interactions, $V=0.2$.}
\label{Fig:Kappa_00_V_theta}
\end{figure}

This changes, however, if we also allow for a moderate
nearest-neighbor interaction $V$ as is shown in
Fig.~\ref{Fig:Kappa_00_V_theta}. Here we have set $U=0$ to concentrate
on the effects of $V$. In contrast to Fig.~\ref{Fig:Kappa_00_U_theta},
where even moderately strong Hubbard interactions $U$ had very little
effect on the charge susceptibility, the nearest-neighbor interaction
dramatically changes the charge response even for strengths as small
as $V=0.2$. For $\theta=0^{\circ}, 11^{\circ}$ the response is
strongly enhanced as compared to the non-interacting case and for
$\theta=20^{\circ}$ there is an instability towards long-range CDW
order. Increasing $V$ further we find a CDW instability for all
$\theta\lesssim 20^\circ$. For even larger canting angles, on the
other hand, there is no charge order as can be seen from the last
column of Fig.~\ref{Fig:Kappa_00_V_theta}.
%
We conclude, that even relatively small nearest-neighbor Coulomb
interactions can lead to CDW instabilities in addition to the magnetic
instabilities discussed earlier. For general $U$ and $V$, we expect a
competition between these different types of orders.

\begin{figure}[t]
	\begin{center}
\hspace{-0.15cm}
\includegraphics[width=0.9 \linewidth]{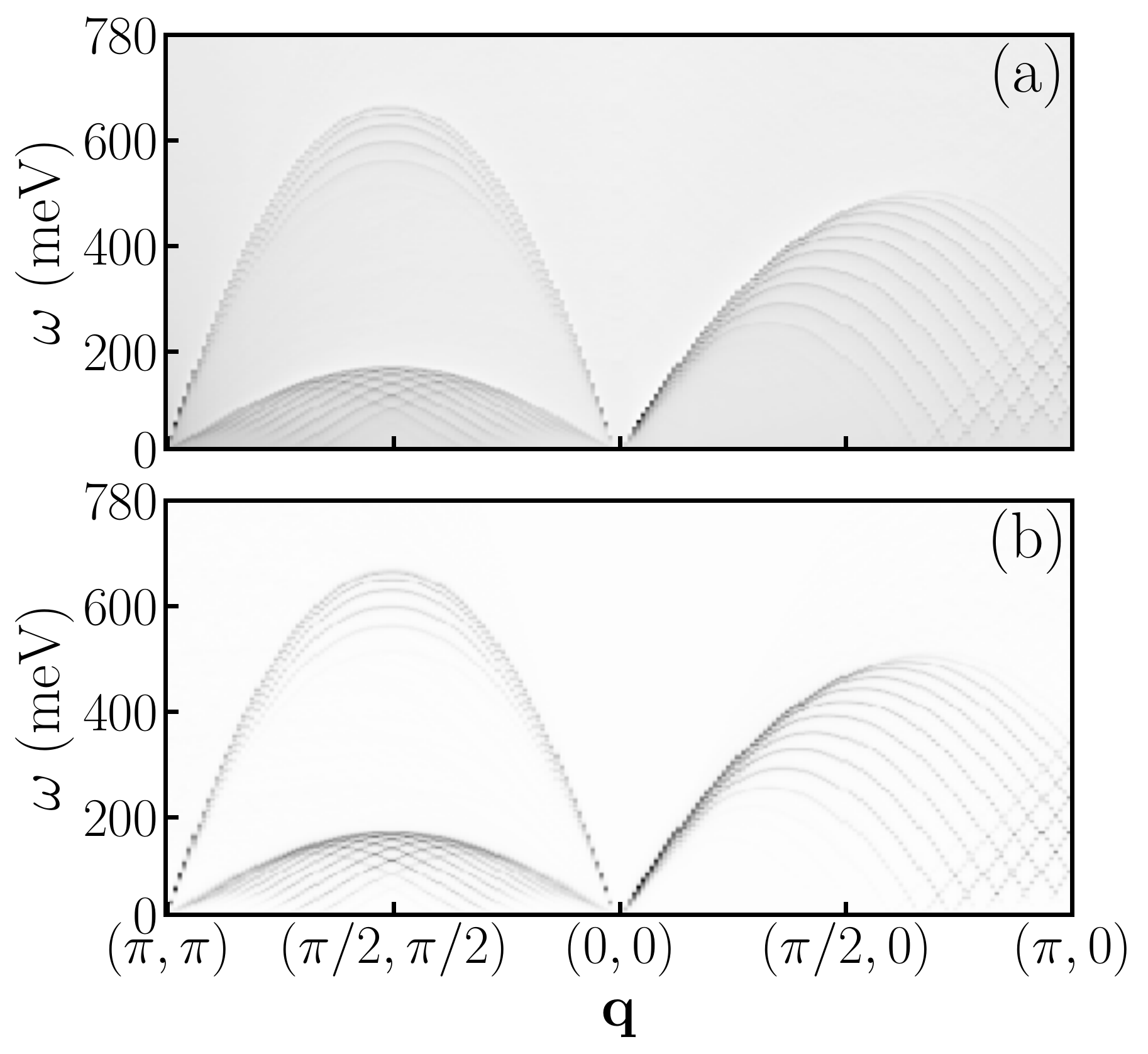}
\vspace{-0.88cm}
\end{center}
\caption{The magnon dispersions along high-symmetry directions for the hole-doped case
with $\langle n \rangle=1.6$, $U=2$, and $V=0$ obtained in RPA. Panel
(a) shows the in-plane magnetic excitations while panel (b) shows the
out-of-plane contributions.}
\label{Fig:Magnon}
\end{figure}

So far, we have concentrated on investigating the electron-doped case
because of its similarities to the hole-doped cuprates. Motivated by a
recent resonant inelastic X-ray scattering (RIXS) study of the
magnetic excitations in Sr$_2$IrO$_4$ in the hole-doped regime, we
have used Eq.~\eqref{Eq:Magnon} to obtain the dynamical spin structure
factors---both in-plane and out-of-plane---within RPA. The results are
shown in Fig.~\ref{Fig:Magnon}. We find several branches of magnetic
excitations with gapless points at $\bq=(0,0)$ and
$\bq=(\pi,\pi)$. The overall structure of the magnetic excitations at
low energies is in good qualitative agreement with the experimental
results reported in Ref.~\cite{BJ_Kim_PRB_2020}. We note, however,
that we also find other branches of magnetic excitations between
$\bq=(0,0)$ and $\bq=(\pi,\pi)$ at much higher energies which have not
been observed experimentally. Since we have not taken into account the
experimental resolution nor the atomic form factors, our results are
meant merely as a qualitative check. A more detailed analysis would
require to reduce our model to an effective Heisenberg or t-J
model. The experimental results in Ref.~\cite{BJ_Kim_PRB_2020} also
suggest that it might be important to also include third neighbor
hopping to accurately describe the dispersions across the full
Brillouin zone, hopping processes which we have neglected here.
%
%

\section{Conclusions}
\label{Sec_Concl}
We have studied a two-dimensional single band Hubbard model with
asymmetric spin-orbit couplings which is a minimal model for the $5d$
layered transition metal oxides such as Sr$_2$IrO$_4$. Most of our
study has been focused on the electron-doped case which is believed
to show similarities to the hole-doped cuprates. As a first step, we
have investigated the electronic band structure of the non-interacting
system as a function of the Ir-O-Ir bond angle. With increasing bond
angle, the spin-dependent hopping terms become more important leading
to a small splitting of the bands. At the same time, the second
nearest-neighbor hopping amplitude decreases and the dispersions
become almost particle-hole symmetric for large canting angles
$\theta\sim 30^\circ$. The almost flat bands which form in the latter
case between high-symmetry points in the Brillouin zone are an
interesting aspect of this model which we plan to further explore in
future studies of the superconducting and possibly topological
properties of this model. To understand the effects of on-site and
nearest-neighbor interactions, we have used exact diagonalizations of
small clusters and the random-phase approximation. From these
calculations a consistent picture of the magnetic instabilities of the
system emerges: For the electron doped case with $\langle n\rangle
=2.4$, $U\gtrsim 2$ and canting angles $\theta\lesssim 15^\circ$
long-range, in-plane antiferromagnetic order forms. This order is
replaced by dominant ferromagnetic fluctuations for larger canting
angles. More generally speaking, we have been able to map out the
regions of the $(U,\theta)$ phase diagram where antiferromagnetic or
ferromagnetic fluctuations dominate. We have found that the borders of
these commensurate magnetic regions around half-filling are defined by
van-Hove singularities with incommensurate fluctuations dominating for
larger doping levels. Furthermore, we have shown that fairly modest
nearest-neighbor Hubbard interactions can give rise to additional CDW
instabilities. The model thus shows an intricate interplay between
various magnetic and charge instabilities as a function of the canting
angle---which controls the strength of the spin-orbit coupling and of
the next-nearest neighbor hopping amplitude---and the on-site and
nearest-neighbor Hubbard interactions $U$ and $V$.

While most of our study has focussed on the electron-doped case, we
have also studied the magnetic excitations on the hole-doped side. We
find that the spin-structure factor has gapless points at $\bq=(0,0)$
and $\bq=(\pi,\pi)$. The magnon dispersions which we obtain in our
model calculations are qualitatively consistent with recent RIXS
experiments on hole-doped Sr$_2$IrO$_4$. While a more accurate
description of the experimentally observed dispersions might require
to also include third neighbor hopping processes, as has been in
suggested in Ref.~\cite{BJ_Kim_PRB_2020}, the minimal model considered
here does already give the right energy scales and overall shape of the
dispersions.

For the future it would be interesting to see whether the model does
support superconductivity and if so, what the dominant superconducting
channels are. This might help to further clarify the similarities and
differences between the electron-doped iridates and the hole-doped
cuprates.

\section*{Acknowledgments}
 We are grateful to Y.~Yanase, G.~Jackeli, B.J.~Kim, I.~Eremin,
 P. Thalmeier, G. Khaliullin, D.K.~Singh and Y.~Bang for fruitful
 discussions.  M.~B. acknowledges S.~A.~Jafari, M.~H.~Zare, and
 R.~Jafari for fruitful discussions.  A.~A. acknowledges the support
 of the Max Planck- POSTECH-Hsinchu Center for Complex Phase
 Materials, and financial support from the National Research
 Foundation (NRF) funded by the Ministry of Science of Korea (Grant
 No. 2016K1A4A01922028). J.~S. acknowledges support by the Natural
 Sciences and Engineering Research Council (NSERC, Canada) and by the
 Deutsche Forschungsgemeinschaft (DFG) via Research Unit FOR 2316.



\bibliography{Refs}

\begin{thebibliography}{70}%
\makeatletter
\providecommand \@ifxundefined [1]{%
 \@ifx{#1\undefined}
}%
\providecommand \@ifnum [1]{%
 \ifnum #1\expandafter \@firstoftwo
 \else \expandafter \@secondoftwo
 \fi
}%
\providecommand \@ifx [1]{%
 \ifx #1\expandafter \@firstoftwo
 \else \expandafter \@secondoftwo
 \fi
}%
\providecommand \natexlab [1]{#1}%
\providecommand \enquote  [1]{``#1''}%
\providecommand \bibnamefont  [1]{#1}%
\providecommand \bibfnamefont [1]{#1}%
\providecommand \citenamefont [1]{#1}%
\providecommand \href@noop [0]{\@secondoftwo}%
\providecommand \href [0]{\begingroup \@sanitize@url \@href}%
\providecommand \@href[1]{\@@startlink{#1}\@@href}%
\providecommand \@@href[1]{\endgroup#1\@@endlink}%
\providecommand \@sanitize@url [0]{\catcode `\\12\catcode `\$12\catcode
  `\&12\catcode `\#12\catcode `\^12\catcode `\_12\catcode `\%12\relax}%
\providecommand \@@startlink[1]{}%
\providecommand \@@endlink[0]{}%
\providecommand \url  [0]{\begingroup\@sanitize@url \@url }%
\providecommand \@url [1]{\endgroup\@href {#1}{\urlprefix }}%
\providecommand \urlprefix  [0]{URL }%
\providecommand \Eprint [0]{\href }%
\providecommand \doibase [0]{http://dx.doi.org/}%
\providecommand \selectlanguage [0]{\@gobble}%
\providecommand \bibinfo  [0]{\@secondoftwo}%
\providecommand \bibfield  [0]{\@secondoftwo}%
\providecommand \translation [1]{[#1]}%
\providecommand \BibitemOpen [0]{}%
\providecommand \bibitemStop [0]{}%
\providecommand \bibitemNoStop [0]{.\EOS\space}%
\providecommand \EOS [0]{\spacefactor3000\relax}%
\providecommand \BibitemShut  [1]{\csname bibitem#1\endcsname}%
\let\auto@bib@innerbib\@empty
\bibitem [{\citenamefont {Dagotto}\ and\ \citenamefont
  {Rice}(1996)}]{Dagotto_Science_1996}%
  \BibitemOpen
  \bibfield  {author} {\bibinfo {author} {\bibfnamefont {E.}~\bibnamefont
  {Dagotto}}\ and\ \bibinfo {author} {\bibfnamefont {T.~M.}\ \bibnamefont
  {Rice}},\ }\href {\doibase 10.1126/science.271.5249.618} {\bibfield
  {journal} {\bibinfo  {journal} {Science}\ }\textbf {\bibinfo {volume}
  {271}},\ \bibinfo {pages} {618} (\bibinfo {year} {1996})}\BibitemShut
  {NoStop}%
\bibitem [{\citenamefont {Tokura}\ and\ \citenamefont
  {Nagaosa}(2000)}]{Nagaosa_Science_2000}%
  \BibitemOpen
  \bibfield  {author} {\bibinfo {author} {\bibfnamefont {Y.}~\bibnamefont
  {Tokura}}\ and\ \bibinfo {author} {\bibfnamefont {N.}~\bibnamefont
  {Nagaosa}},\ }\href {\doibase 10.1126/science.288.5465.462} {\bibfield
  {journal} {\bibinfo  {journal} {Science}\ }\textbf {\bibinfo {volume}
  {288}},\ \bibinfo {pages} {462} (\bibinfo {year} {2000})}\BibitemShut
  {NoStop}%
\bibitem [{\citenamefont {Dagotto}(2005)}]{Dagotto_Science_2005}%
  \BibitemOpen
  \bibfield  {author} {\bibinfo {author} {\bibfnamefont {E.}~\bibnamefont
  {Dagotto}},\ }\href {\doibase 10.1126/science.1107559} {\bibfield  {journal}
  {\bibinfo  {journal} {Science}\ }\textbf {\bibinfo {volume} {309}},\ \bibinfo
  {pages} {257} (\bibinfo {year} {2005})}\BibitemShut {NoStop}%
\bibitem [{\citenamefont {Lee}\ \emph {et~al.}(2006)\citenamefont {Lee},
  \citenamefont {Nagaosa},\ and\ \citenamefont {Wen}}]{Wen_Rev_Mod_Phys_2006}%
  \BibitemOpen
  \bibfield  {author} {\bibinfo {author} {\bibfnamefont {P.~A.}\ \bibnamefont
  {Lee}}, \bibinfo {author} {\bibfnamefont {N.}~\bibnamefont {Nagaosa}}, \ and\
  \bibinfo {author} {\bibfnamefont {X.-G.}\ \bibnamefont {Wen}},\ }\href
  {\doibase 10.1103/RevModPhys.78.17} {\bibfield  {journal} {\bibinfo
  {journal} {Rev. Mod. Phys.}\ }\textbf {\bibinfo {volume} {78}},\ \bibinfo
  {pages} {17} (\bibinfo {year} {2006})}\BibitemShut {NoStop}%
\bibitem [{\citenamefont {{Lake}}\ \emph {et~al.}(2010)\citenamefont {{Lake}},
  \citenamefont {{Tsvelik}}, \citenamefont {{Notbohm}}, \citenamefont {{Alan
  Tennant}}, \citenamefont {{Perring}}, \citenamefont {{Reehuis}},
  \citenamefont {{Sekar}}, \citenamefont {{Krabbes}},\ and\ \citenamefont
  {{B{\"u}chner}}}]{Buchner_NatPhys_2009}%
  \BibitemOpen
  \bibfield  {author} {\bibinfo {author} {\bibfnamefont {B.}~\bibnamefont
  {{Lake}}}, \bibinfo {author} {\bibfnamefont {A.~M.}\ \bibnamefont
  {{Tsvelik}}}, \bibinfo {author} {\bibfnamefont {S.}~\bibnamefont
  {{Notbohm}}}, \bibinfo {author} {\bibfnamefont {D.}~\bibnamefont {{Alan
  Tennant}}}, \bibinfo {author} {\bibfnamefont {T.~G.}\ \bibnamefont
  {{Perring}}}, \bibinfo {author} {\bibfnamefont {M.}~\bibnamefont
  {{Reehuis}}}, \bibinfo {author} {\bibfnamefont {C.}~\bibnamefont {{Sekar}}},
  \bibinfo {author} {\bibfnamefont {G.}~\bibnamefont {{Krabbes}}}, \ and\
  \bibinfo {author} {\bibfnamefont {B.}~\bibnamefont {{B{\"u}chner}}},\ }\href
  {\doibase 10.1038/nphys1462} {\bibfield  {journal} {\bibinfo  {journal}
  {Nature Physics}\ }\textbf {\bibinfo {volume} {6}},\ \bibinfo {pages} {50}
  (\bibinfo {year} {2010})}\BibitemShut {NoStop}%
\bibitem [{\citenamefont {Scalapino}(2012)}]{Scalapino_Rev_Mod_Phys_2012}%
  \BibitemOpen
  \bibfield  {author} {\bibinfo {author} {\bibfnamefont {D.~J.}\ \bibnamefont
  {Scalapino}},\ }\href {\doibase 10.1103/RevModPhys.84.1383} {\bibfield
  {journal} {\bibinfo  {journal} {Rev. Mod. Phys.}\ }\textbf {\bibinfo {volume}
  {84}},\ \bibinfo {pages} {1383} (\bibinfo {year} {2012})}\BibitemShut
  {NoStop}%
\bibitem [{\citenamefont {Choi}\ \emph {et~al.}(2012)\citenamefont {Choi},
  \citenamefont {Coldea}, \citenamefont {Kolmogorov}, \citenamefont
  {Lancaster}, \citenamefont {Mazin}, \citenamefont {Blundell}, \citenamefont
  {Radaelli}, \citenamefont {Singh}, \citenamefont {Gegenwart}, \citenamefont
  {Choi}, \citenamefont {Cheong}, \citenamefont {Baker}, \citenamefont
  {Stock},\ and\ \citenamefont {Taylor}}]{Taylor_PRL_2012}%
  \BibitemOpen
  \bibfield  {author} {\bibinfo {author} {\bibfnamefont {S.~K.}\ \bibnamefont
  {Choi}}, \bibinfo {author} {\bibfnamefont {R.}~\bibnamefont {Coldea}},
  \bibinfo {author} {\bibfnamefont {A.~N.}\ \bibnamefont {Kolmogorov}},
  \bibinfo {author} {\bibfnamefont {T.}~\bibnamefont {Lancaster}}, \bibinfo
  {author} {\bibfnamefont {I.~I.}\ \bibnamefont {Mazin}}, \bibinfo {author}
  {\bibfnamefont {S.~J.}\ \bibnamefont {Blundell}}, \bibinfo {author}
  {\bibfnamefont {P.~G.}\ \bibnamefont {Radaelli}}, \bibinfo {author}
  {\bibfnamefont {Y.}~\bibnamefont {Singh}}, \bibinfo {author} {\bibfnamefont
  {P.}~\bibnamefont {Gegenwart}}, \bibinfo {author} {\bibfnamefont {K.~R.}\
  \bibnamefont {Choi}}, \bibinfo {author} {\bibfnamefont {S.-W.}\ \bibnamefont
  {Cheong}}, \bibinfo {author} {\bibfnamefont {P.~J.}\ \bibnamefont {Baker}},
  \bibinfo {author} {\bibfnamefont {C.}~\bibnamefont {Stock}}, \ and\ \bibinfo
  {author} {\bibfnamefont {J.}~\bibnamefont {Taylor}},\ }\href {\doibase
  10.1103/PhysRevLett.108.127204} {\bibfield  {journal} {\bibinfo  {journal}
  {Phys. Rev. Lett.}\ }\textbf {\bibinfo {volume} {108}},\ \bibinfo {pages}
  {127204} (\bibinfo {year} {2012})}\BibitemShut {NoStop}%
\bibitem [{\citenamefont {{Chun}}\ \emph {et~al.}(2015)\citenamefont {{Chun}},
  \citenamefont {{Kim}}, \citenamefont {{Kim}}, \citenamefont {{Zheng}},
  \citenamefont {{Stoumpos}}, \citenamefont {{Malliakas}}, \citenamefont
  {{Mitchell}}, \citenamefont {{Mehlawat}}, \citenamefont {{Singh}},
  \citenamefont {{Choi}}, \citenamefont {{Gog}}, \citenamefont {{Al-Zein}},
  \citenamefont {{Moretti Sala}}, \citenamefont {{Krisch}}, \citenamefont
  {{Chaloupka}}, \citenamefont {{Jackeli}}, \citenamefont {{Khaliullin}},\ and\
  \citenamefont {{Kim}}}]{BJ_Kim_NatPhys_2015}%
  \BibitemOpen
  \bibfield  {author} {\bibinfo {author} {\bibfnamefont {S.~H.}\ \bibnamefont
  {{Chun}}}, \bibinfo {author} {\bibfnamefont {J.-W.}\ \bibnamefont {{Kim}}},
  \bibinfo {author} {\bibfnamefont {J.}~\bibnamefont {{Kim}}}, \bibinfo
  {author} {\bibfnamefont {H.}~\bibnamefont {{Zheng}}}, \bibinfo {author}
  {\bibfnamefont {C.~C.}\ \bibnamefont {{Stoumpos}}}, \bibinfo {author}
  {\bibfnamefont {C.~D.}\ \bibnamefont {{Malliakas}}}, \bibinfo {author}
  {\bibfnamefont {J.~F.}\ \bibnamefont {{Mitchell}}}, \bibinfo {author}
  {\bibfnamefont {K.}~\bibnamefont {{Mehlawat}}}, \bibinfo {author}
  {\bibfnamefont {Y.}~\bibnamefont {{Singh}}}, \bibinfo {author} {\bibfnamefont
  {Y.}~\bibnamefont {{Choi}}}, \bibinfo {author} {\bibfnamefont
  {T.}~\bibnamefont {{Gog}}}, \bibinfo {author} {\bibfnamefont
  {A.}~\bibnamefont {{Al-Zein}}}, \bibinfo {author} {\bibfnamefont
  {M.}~\bibnamefont {{Moretti Sala}}}, \bibinfo {author} {\bibfnamefont
  {M.}~\bibnamefont {{Krisch}}}, \bibinfo {author} {\bibfnamefont
  {J.}~\bibnamefont {{Chaloupka}}}, \bibinfo {author} {\bibfnamefont
  {G.}~\bibnamefont {{Jackeli}}}, \bibinfo {author} {\bibfnamefont
  {G.}~\bibnamefont {{Khaliullin}}}, \ and\ \bibinfo {author} {\bibfnamefont
  {B.~J.}\ \bibnamefont {{Kim}}},\ }\href@noop {} {\bibfield  {journal}
  {\bibinfo  {journal} {arXiv e-prints}\ ,\ \bibinfo {pages}
  {arXiv:1504.03618}} (\bibinfo {year} {2015})},\ \Eprint
  {http://arxiv.org/abs/1504.03618} {1504.03618 [cond-mat.mtrl-sci]}
  \BibitemShut {NoStop}%
\bibitem [{\citenamefont {Cui}\ \emph {et~al.}(2016)\citenamefont {Cui},
  \citenamefont {Cheng}, \citenamefont {Fan}, \citenamefont {Taylor},
  \citenamefont {Calder}, \citenamefont {McGuire}, \citenamefont {Yan},
  \citenamefont {Meyers}, \citenamefont {Li}, \citenamefont {Cai},
  \citenamefont {Jiao}, \citenamefont {Choi}, \citenamefont {Haskel},
  \citenamefont {Gotou}, \citenamefont {Uwatoko}, \citenamefont {Chakhalian},
  \citenamefont {Christianson}, \citenamefont {Yunoki}, \citenamefont
  {Goodenough},\ and\ \citenamefont {Zhou}}]{Zhou_PRL_2016}%
  \BibitemOpen
  \bibfield  {author} {\bibinfo {author} {\bibfnamefont {Q.}~\bibnamefont
  {Cui}}, \bibinfo {author} {\bibfnamefont {J.-G.}\ \bibnamefont {Cheng}},
  \bibinfo {author} {\bibfnamefont {W.}~\bibnamefont {Fan}}, \bibinfo {author}
  {\bibfnamefont {A.~E.}\ \bibnamefont {Taylor}}, \bibinfo {author}
  {\bibfnamefont {S.}~\bibnamefont {Calder}}, \bibinfo {author} {\bibfnamefont
  {M.~A.}\ \bibnamefont {McGuire}}, \bibinfo {author} {\bibfnamefont {J.-Q.}\
  \bibnamefont {Yan}}, \bibinfo {author} {\bibfnamefont {D.}~\bibnamefont
  {Meyers}}, \bibinfo {author} {\bibfnamefont {X.}~\bibnamefont {Li}}, \bibinfo
  {author} {\bibfnamefont {Y.~Q.}\ \bibnamefont {Cai}}, \bibinfo {author}
  {\bibfnamefont {Y.~Y.}\ \bibnamefont {Jiao}}, \bibinfo {author}
  {\bibfnamefont {Y.}~\bibnamefont {Choi}}, \bibinfo {author} {\bibfnamefont
  {D.}~\bibnamefont {Haskel}}, \bibinfo {author} {\bibfnamefont
  {H.}~\bibnamefont {Gotou}}, \bibinfo {author} {\bibfnamefont
  {Y.}~\bibnamefont {Uwatoko}}, \bibinfo {author} {\bibfnamefont
  {J.}~\bibnamefont {Chakhalian}}, \bibinfo {author} {\bibfnamefont {A.~D.}\
  \bibnamefont {Christianson}}, \bibinfo {author} {\bibfnamefont
  {S.}~\bibnamefont {Yunoki}}, \bibinfo {author} {\bibfnamefont {J.~B.}\
  \bibnamefont {Goodenough}}, \ and\ \bibinfo {author} {\bibfnamefont {J.-S.}\
  \bibnamefont {Zhou}},\ }\href {\doibase 10.1103/PhysRevLett.117.176603}
  {\bibfield  {journal} {\bibinfo  {journal} {Phys. Rev. Lett.}\ }\textbf
  {\bibinfo {volume} {117}},\ \bibinfo {pages} {176603} (\bibinfo {year}
  {2016})}\BibitemShut {NoStop}%
\bibitem [{\citenamefont {Majumder}\ \emph {et~al.}(2018)\citenamefont
  {Majumder}, \citenamefont {Manna}, \citenamefont {Simutis}, \citenamefont
  {Orain}, \citenamefont {Dey}, \citenamefont {Freund}, \citenamefont {Jesche},
  \citenamefont {Khasanov}, \citenamefont {Biswas}, \citenamefont {Bykova},
  \citenamefont {Dubrovinskaia}, \citenamefont {Dubrovinsky}, \citenamefont
  {Yadav}, \citenamefont {Hozoi}, \citenamefont {Nishimoto}, \citenamefont
  {Tsirlin},\ and\ \citenamefont {Gegenwart}}]{Majumdar_PRL_2018}%
  \BibitemOpen
  \bibfield  {author} {\bibinfo {author} {\bibfnamefont {M.}~\bibnamefont
  {Majumder}}, \bibinfo {author} {\bibfnamefont {R.~S.}\ \bibnamefont {Manna}},
  \bibinfo {author} {\bibfnamefont {G.}~\bibnamefont {Simutis}}, \bibinfo
  {author} {\bibfnamefont {J.~C.}\ \bibnamefont {Orain}}, \bibinfo {author}
  {\bibfnamefont {T.}~\bibnamefont {Dey}}, \bibinfo {author} {\bibfnamefont
  {F.}~\bibnamefont {Freund}}, \bibinfo {author} {\bibfnamefont
  {A.}~\bibnamefont {Jesche}}, \bibinfo {author} {\bibfnamefont
  {R.}~\bibnamefont {Khasanov}}, \bibinfo {author} {\bibfnamefont {P.~K.}\
  \bibnamefont {Biswas}}, \bibinfo {author} {\bibfnamefont {E.}~\bibnamefont
  {Bykova}}, \bibinfo {author} {\bibfnamefont {N.}~\bibnamefont
  {Dubrovinskaia}}, \bibinfo {author} {\bibfnamefont {L.~S.}\ \bibnamefont
  {Dubrovinsky}}, \bibinfo {author} {\bibfnamefont {R.}~\bibnamefont {Yadav}},
  \bibinfo {author} {\bibfnamefont {L.}~\bibnamefont {Hozoi}}, \bibinfo
  {author} {\bibfnamefont {S.}~\bibnamefont {Nishimoto}}, \bibinfo {author}
  {\bibfnamefont {A.~A.}\ \bibnamefont {Tsirlin}}, \ and\ \bibinfo {author}
  {\bibfnamefont {P.}~\bibnamefont {Gegenwart}},\ }\href {\doibase
  10.1103/PhysRevLett.120.237202} {\bibfield  {journal} {\bibinfo  {journal}
  {Phys. Rev. Lett.}\ }\textbf {\bibinfo {volume} {120}},\ \bibinfo {pages}
  {237202} (\bibinfo {year} {2018})}\BibitemShut {NoStop}%
\bibitem [{\citenamefont {Kitagawa}\ \emph {et~al.}(2018)\citenamefont
  {Kitagawa}, \citenamefont {Takayama}, \citenamefont {Matsumoto},
  \citenamefont {Kato}, \citenamefont {Takano}, \citenamefont {Kishimoto},
  \citenamefont {Bette}, \citenamefont {Dinnebier}, \citenamefont {Jackeli},\
  and\ \citenamefont {Takagi}}]{Kitagawa_Nature_2018}%
  \BibitemOpen
  \bibfield  {author} {\bibinfo {author} {\bibfnamefont {K.}~\bibnamefont
  {Kitagawa}}, \bibinfo {author} {\bibfnamefont {T.}~\bibnamefont {Takayama}},
  \bibinfo {author} {\bibfnamefont {Y.}~\bibnamefont {Matsumoto}}, \bibinfo
  {author} {\bibfnamefont {A.}~\bibnamefont {Kato}}, \bibinfo {author}
  {\bibfnamefont {R.}~\bibnamefont {Takano}}, \bibinfo {author} {\bibfnamefont
  {Y.}~\bibnamefont {Kishimoto}}, \bibinfo {author} {\bibfnamefont
  {S.}~\bibnamefont {Bette}}, \bibinfo {author} {\bibfnamefont
  {R.}~\bibnamefont {Dinnebier}}, \bibinfo {author} {\bibfnamefont
  {G.}~\bibnamefont {Jackeli}}, \ and\ \bibinfo {author} {\bibfnamefont
  {H.}~\bibnamefont {Takagi}},\ }\href {\doibase 10.1038/nature25482}
  {\bibfield  {journal} {\bibinfo  {journal} {Nature}\ }\textbf {\bibinfo
  {volume} {554}},\ \bibinfo {pages} {341} (\bibinfo {year}
  {2018})}\BibitemShut {NoStop}%
\bibitem [{\citenamefont {Bertinshaw}\ \emph {et~al.}(2019)\citenamefont
  {Bertinshaw}, \citenamefont {Kim}, \citenamefont {Khaliullin},\ and\
  \citenamefont {Kim}}]{BJ_Kim_Annual_2019}%
  \BibitemOpen
  \bibfield  {author} {\bibinfo {author} {\bibfnamefont {J.}~\bibnamefont
  {Bertinshaw}}, \bibinfo {author} {\bibfnamefont {Y.}~\bibnamefont {Kim}},
  \bibinfo {author} {\bibfnamefont {G.}~\bibnamefont {Khaliullin}}, \ and\
  \bibinfo {author} {\bibfnamefont {B.}~\bibnamefont {Kim}},\ }\href {\doibase
  10.1146/annurev-conmatphys-031218-013113} {\bibfield  {journal} {\bibinfo
  {journal} {Annual Review of Condensed Matter Physics}\ }\textbf {\bibinfo
  {volume} {10}},\ \bibinfo {pages} {315} (\bibinfo {year} {2019})}\BibitemShut
  {NoStop}%
\bibitem [{\citenamefont {Sun}\ \emph {et~al.}(2020)\citenamefont {Sun},
  \citenamefont {Zhu},\ and\ \citenamefont {Weng}}]{Weng_PhysRevRes_2020}%
  \BibitemOpen
  \bibfield  {author} {\bibinfo {author} {\bibfnamefont {R.-Y.}\ \bibnamefont
  {Sun}}, \bibinfo {author} {\bibfnamefont {Z.}~\bibnamefont {Zhu}}, \ and\
  \bibinfo {author} {\bibfnamefont {Z.-Y.}\ \bibnamefont {Weng}},\ }\href
  {\doibase 10.1103/PhysRevResearch.2.033007} {\bibfield  {journal} {\bibinfo
  {journal} {Phys. Rev. Research}\ }\textbf {\bibinfo {volume} {2}},\ \bibinfo
  {pages} {033007} (\bibinfo {year} {2020})}\BibitemShut {NoStop}%
\bibitem [{\citenamefont {{Tranquada}}\ \emph {et~al.}(1995)\citenamefont
  {{Tranquada}}, \citenamefont {{Sternlieb}}, \citenamefont {{Axe}},
  \citenamefont {{Nakamura}},\ and\ \citenamefont
  {{Uchida}}}]{Uchida_Nature_1995}%
  \BibitemOpen
  \bibfield  {author} {\bibinfo {author} {\bibfnamefont {J.~M.}\ \bibnamefont
  {{Tranquada}}}, \bibinfo {author} {\bibfnamefont {B.~J.}\ \bibnamefont
  {{Sternlieb}}}, \bibinfo {author} {\bibfnamefont {J.~D.}\ \bibnamefont
  {{Axe}}}, \bibinfo {author} {\bibfnamefont {Y.}~\bibnamefont {{Nakamura}}}, \
  and\ \bibinfo {author} {\bibfnamefont {S.}~\bibnamefont {{Uchida}}},\ }\href
  {\doibase 10.1038/375561a0} {\bibfield  {journal} {\bibinfo  {journal}
  {\nat}\ }\textbf {\bibinfo {volume} {375}},\ \bibinfo {pages} {561} (\bibinfo
  {year} {1995})}\BibitemShut {NoStop}%
\bibitem [{\citenamefont {Mook}\ \emph {et~al.}(2002)\citenamefont {Mook},
  \citenamefont {Dai},\ and\ \citenamefont {{Do\ifmmode \breve{g}\else {\u
  g}\fi{}an}}}]{Dogan_PRL_2002}%
  \BibitemOpen
  \bibfield  {author} {\bibinfo {author} {\bibfnamefont {H.~A.}\ \bibnamefont
  {Mook}}, \bibinfo {author} {\bibfnamefont {P.}~\bibnamefont {Dai}}, \ and\
  \bibinfo {author} {\bibfnamefont {F.}~\bibnamefont {{Do\ifmmode
  \breve{g}\else {\u g}\fi{}an}}},\ }\href {\doibase
  10.1103/PhysRevLett.88.097004} {\bibfield  {journal} {\bibinfo  {journal}
  {Phys. Rev. Lett.}\ }\textbf {\bibinfo {volume} {88}},\ \bibinfo {pages}
  {097004} (\bibinfo {year} {2002})}\BibitemShut {NoStop}%
\bibitem [{\citenamefont {Kajimoto}\ \emph {et~al.}(2003)\citenamefont
  {Kajimoto}, \citenamefont {Ishizaka}, \citenamefont {Yoshizawa},\ and\
  \citenamefont {Tokura}}]{Kajimoto_PRB_2003}%
  \BibitemOpen
  \bibfield  {author} {\bibinfo {author} {\bibfnamefont {R.}~\bibnamefont
  {Kajimoto}}, \bibinfo {author} {\bibfnamefont {K.}~\bibnamefont {Ishizaka}},
  \bibinfo {author} {\bibfnamefont {H.}~\bibnamefont {Yoshizawa}}, \ and\
  \bibinfo {author} {\bibfnamefont {Y.}~\bibnamefont {Tokura}},\ }\href
  {\doibase 10.1103/PhysRevB.67.014511} {\bibfield  {journal} {\bibinfo
  {journal} {Phys. Rev. B}\ }\textbf {\bibinfo {volume} {67}},\ \bibinfo
  {pages} {014511} (\bibinfo {year} {2003})}\BibitemShut {NoStop}%
\bibitem [{\citenamefont {Howald}\ \emph {et~al.}(2003)\citenamefont {Howald},
  \citenamefont {Eisaki}, \citenamefont {Kaneko}, \citenamefont {Greven},\ and\
  \citenamefont {Kapitulnik}}]{Kapitolnic_PRB_2003}%
  \BibitemOpen
  \bibfield  {author} {\bibinfo {author} {\bibfnamefont {C.}~\bibnamefont
  {Howald}}, \bibinfo {author} {\bibfnamefont {H.}~\bibnamefont {Eisaki}},
  \bibinfo {author} {\bibfnamefont {N.}~\bibnamefont {Kaneko}}, \bibinfo
  {author} {\bibfnamefont {M.}~\bibnamefont {Greven}}, \ and\ \bibinfo {author}
  {\bibfnamefont {A.}~\bibnamefont {Kapitulnik}},\ }\href {\doibase
  10.1103/PhysRevB.67.014533} {\bibfield  {journal} {\bibinfo  {journal} {Phys.
  Rev. B}\ }\textbf {\bibinfo {volume} {67}},\ \bibinfo {pages} {014533}
  (\bibinfo {year} {2003})}\BibitemShut {NoStop}%
\bibitem [{\citenamefont {{Chen}}\ \emph {et~al.}(2018)\citenamefont {{Chen}},
  \citenamefont {{Schmehr}}, \citenamefont {{Islam}}, \citenamefont {{Porter}},
  \citenamefont {{Zoghlin}}, \citenamefont {{Finkelstein}}, \citenamefont
  {{Ruff}},\ and\ \citenamefont {{Wilson}}}]{Wilson_Nat_Comm_2018}%
  \BibitemOpen
  \bibfield  {author} {\bibinfo {author} {\bibfnamefont {X.}~\bibnamefont
  {{Chen}}}, \bibinfo {author} {\bibfnamefont {J.~L.}\ \bibnamefont
  {{Schmehr}}}, \bibinfo {author} {\bibfnamefont {Z.}~\bibnamefont {{Islam}}},
  \bibinfo {author} {\bibfnamefont {Z.}~\bibnamefont {{Porter}}}, \bibinfo
  {author} {\bibfnamefont {E.}~\bibnamefont {{Zoghlin}}}, \bibinfo {author}
  {\bibfnamefont {K.}~\bibnamefont {{Finkelstein}}}, \bibinfo {author}
  {\bibfnamefont {J.~P.~C.}\ \bibnamefont {{Ruff}}}, \ and\ \bibinfo {author}
  {\bibfnamefont {S.~D.}\ \bibnamefont {{Wilson}}},\ }\href {\doibase
  10.1038/s41467-017-02647-1} {\bibfield  {journal} {\bibinfo  {journal}
  {Nature Communications}\ }\textbf {\bibinfo {volume} {9}},\ \bibinfo {eid}
  {103} (\bibinfo {year} {2018})}\BibitemShut {NoStop}%
\bibitem [{\citenamefont {{Pesin}}\ and\ \citenamefont
  {{Balents}}(2010)}]{Pesin_Nature_2010}%
  \BibitemOpen
  \bibfield  {author} {\bibinfo {author} {\bibfnamefont {D.}~\bibnamefont
  {{Pesin}}}\ and\ \bibinfo {author} {\bibfnamefont {L.}~\bibnamefont
  {{Balents}}},\ }\href {\doibase 10.1038/nphys1606} {\bibfield  {journal}
  {\bibinfo  {journal} {Nature Physics}\ }\textbf {\bibinfo {volume} {6}},\
  \bibinfo {pages} {376} (\bibinfo {year} {2010})}\BibitemShut {NoStop}%
\bibitem [{\citenamefont {{Witczak-Krempa}}\ \emph {et~al.}(2014)\citenamefont
  {{Witczak-Krempa}}, \citenamefont {{Chen}}, \citenamefont {{Kim}},\ and\
  \citenamefont {{Balents}}}]{Witzcak_Krampa_Annual_2015}%
  \BibitemOpen
  \bibfield  {author} {\bibinfo {author} {\bibfnamefont {W.}~\bibnamefont
  {{Witczak-Krempa}}}, \bibinfo {author} {\bibfnamefont {G.}~\bibnamefont
  {{Chen}}}, \bibinfo {author} {\bibfnamefont {Y.~B.}\ \bibnamefont {{Kim}}}, \
  and\ \bibinfo {author} {\bibfnamefont {L.}~\bibnamefont {{Balents}}},\ }\href
  {\doibase 10.1146/annurev-conmatphys-020911-125138} {\bibfield  {journal}
  {\bibinfo  {journal} {Annual Review of Condensed Matter Physics}\ }\textbf
  {\bibinfo {volume} {5}},\ \bibinfo {pages} {57} (\bibinfo {year}
  {2014})}\BibitemShut {NoStop}%
\bibitem [{\citenamefont {{Rau}}\ \emph {et~al.}(2016)\citenamefont {{Rau}},
  \citenamefont {{Lee}},\ and\ \citenamefont {{Kee}}}]{Kee_Annual_2016}%
  \BibitemOpen
  \bibfield  {author} {\bibinfo {author} {\bibfnamefont {J.~G.}\ \bibnamefont
  {{Rau}}}, \bibinfo {author} {\bibfnamefont {E.~K.-H.}\ \bibnamefont {{Lee}}},
  \ and\ \bibinfo {author} {\bibfnamefont {H.-Y.}\ \bibnamefont {{Kee}}},\
  }\href {\doibase 10.1146/annurev-conmatphys-031115-011319} {\bibfield
  {journal} {\bibinfo  {journal} {Annual Review of Condensed Matter Physics}\
  }\textbf {\bibinfo {volume} {7}},\ \bibinfo {pages} {195} (\bibinfo {year}
  {2016})}\BibitemShut {NoStop}%
\bibitem [{\citenamefont {Fujiyama}\ \emph {et~al.}(2012)\citenamefont
  {Fujiyama}, \citenamefont {Ohsumi}, \citenamefont {Komesu}, \citenamefont
  {Matsuno}, \citenamefont {Kim}, \citenamefont {Takata}, \citenamefont
  {Arima},\ and\ \citenamefont {Takagi}}]{Fujiyama_PRL_2012}%
  \BibitemOpen
  \bibfield  {author} {\bibinfo {author} {\bibfnamefont {S.}~\bibnamefont
  {Fujiyama}}, \bibinfo {author} {\bibfnamefont {H.}~\bibnamefont {Ohsumi}},
  \bibinfo {author} {\bibfnamefont {T.}~\bibnamefont {Komesu}}, \bibinfo
  {author} {\bibfnamefont {J.}~\bibnamefont {Matsuno}}, \bibinfo {author}
  {\bibfnamefont {B.~J.}\ \bibnamefont {Kim}}, \bibinfo {author} {\bibfnamefont
  {M.}~\bibnamefont {Takata}}, \bibinfo {author} {\bibfnamefont
  {T.}~\bibnamefont {Arima}}, \ and\ \bibinfo {author} {\bibfnamefont
  {H.}~\bibnamefont {Takagi}},\ }\href {\doibase
  10.1103/PhysRevLett.108.247212} {\bibfield  {journal} {\bibinfo  {journal}
  {Phys. Rev. Lett.}\ }\textbf {\bibinfo {volume} {108}},\ \bibinfo {pages}
  {247212} (\bibinfo {year} {2012})}\BibitemShut {NoStop}%
\bibitem [{\citenamefont {Wang}\ and\ \citenamefont
  {Senthil}(2011{\natexlab{a}})}]{Wang_PRL_2011}%
  \BibitemOpen
  \bibfield  {author} {\bibinfo {author} {\bibfnamefont {F.}~\bibnamefont
  {Wang}}\ and\ \bibinfo {author} {\bibfnamefont {T.}~\bibnamefont {Senthil}},\
  }\href {\doibase 10.1103/PhysRevLett.106.136402} {\bibfield  {journal}
  {\bibinfo  {journal} {Phys. Rev. Lett.}\ }\textbf {\bibinfo {volume} {106}},\
  \bibinfo {pages} {136402} (\bibinfo {year} {2011}{\natexlab{a}})}\BibitemShut
  {NoStop}%
\bibitem [{\citenamefont {Kim}\ \emph {et~al.}(2014)\citenamefont {Kim},
  \citenamefont {Krupin}, \citenamefont {Denlinger}, \citenamefont {Bostwick},
  \citenamefont {Rotenberg}, \citenamefont {Zhao}, \citenamefont {Mitchell},
  \citenamefont {Allen},\ and\ \citenamefont {Kim}}]{BJKim_Science_2014}%
  \BibitemOpen
  \bibfield  {author} {\bibinfo {author} {\bibfnamefont {Y.~K.}\ \bibnamefont
  {Kim}}, \bibinfo {author} {\bibfnamefont {O.}~\bibnamefont {Krupin}},
  \bibinfo {author} {\bibfnamefont {J.~D.}\ \bibnamefont {Denlinger}}, \bibinfo
  {author} {\bibfnamefont {A.}~\bibnamefont {Bostwick}}, \bibinfo {author}
  {\bibfnamefont {E.}~\bibnamefont {Rotenberg}}, \bibinfo {author}
  {\bibfnamefont {Q.}~\bibnamefont {Zhao}}, \bibinfo {author} {\bibfnamefont
  {J.~F.}\ \bibnamefont {Mitchell}}, \bibinfo {author} {\bibfnamefont {J.~W.}\
  \bibnamefont {Allen}}, \ and\ \bibinfo {author} {\bibfnamefont {B.~J.}\
  \bibnamefont {Kim}},\ }\href {\doibase 10.1126/science.1251151} {\bibfield
  {journal} {\bibinfo  {journal} {Science}\ }\textbf {\bibinfo {volume}
  {345}},\ \bibinfo {pages} {187} (\bibinfo {year} {2014})}\BibitemShut
  {NoStop}%
\bibitem [{\citenamefont {Watanabe}\ \emph {et~al.}(2013)\citenamefont
  {Watanabe}, \citenamefont {Shirakawa},\ and\ \citenamefont
  {Yunoki}}]{Watanabe_PRL_2013}%
  \BibitemOpen
  \bibfield  {author} {\bibinfo {author} {\bibfnamefont {H.}~\bibnamefont
  {Watanabe}}, \bibinfo {author} {\bibfnamefont {T.}~\bibnamefont {Shirakawa}},
  \ and\ \bibinfo {author} {\bibfnamefont {S.}~\bibnamefont {Yunoki}},\ }\href
  {\doibase 10.1103/PhysRevLett.110.027002} {\bibfield  {journal} {\bibinfo
  {journal} {Phys. Rev. Lett.}\ }\textbf {\bibinfo {volume} {110}},\ \bibinfo
  {pages} {027002} (\bibinfo {year} {2013})}\BibitemShut {NoStop}%
\bibitem [{\citenamefont {Meng}\ \emph {et~al.}(2014)\citenamefont {Meng},
  \citenamefont {Kim},\ and\ \citenamefont {Kee}}]{Meng_PRL_2104}%
  \BibitemOpen
  \bibfield  {author} {\bibinfo {author} {\bibfnamefont {Z.~Y.}\ \bibnamefont
  {Meng}}, \bibinfo {author} {\bibfnamefont {Y.~B.}\ \bibnamefont {Kim}}, \
  and\ \bibinfo {author} {\bibfnamefont {H.-Y.}\ \bibnamefont {Kee}},\ }\href
  {\doibase 10.1103/PhysRevLett.113.177003} {\bibfield  {journal} {\bibinfo
  {journal} {Phys. Rev. Lett.}\ }\textbf {\bibinfo {volume} {113}},\ \bibinfo
  {pages} {177003} (\bibinfo {year} {2014})}\BibitemShut {NoStop}%
\bibitem [{\citenamefont {de~la Torre}\ \emph {et~al.}(2015)\citenamefont
  {de~la Torre}, \citenamefont {{McKeown Walker}}, \citenamefont {Bruno},
  \citenamefont {Ricc{\'o}}, \citenamefont {Wang}, \citenamefont {{Gutierrez
  Lezama}}, \citenamefont {Scheerer}, \citenamefont {Giriat}, \citenamefont
  {Jaccard}, \citenamefont {Berthod}, \citenamefont {Kim}, \citenamefont
  {Hoesch}, \citenamefont {Hunter}, \citenamefont {Perry}, \citenamefont
  {Tamai},\ and\ \citenamefont {Baumberger}}]{delaToore_PRL_2015}%
  \BibitemOpen
  \bibfield  {author} {\bibinfo {author} {\bibfnamefont {A.}~\bibnamefont
  {de~la Torre}}, \bibinfo {author} {\bibfnamefont {S.}~\bibnamefont {{McKeown
  Walker}}}, \bibinfo {author} {\bibfnamefont {F.~Y.}\ \bibnamefont {Bruno}},
  \bibinfo {author} {\bibfnamefont {S.}~\bibnamefont {Ricc{\'o}}}, \bibinfo
  {author} {\bibfnamefont {Z.}~\bibnamefont {Wang}}, \bibinfo {author}
  {\bibfnamefont {I.}~\bibnamefont {{Gutierrez Lezama}}}, \bibinfo {author}
  {\bibfnamefont {G.}~\bibnamefont {Scheerer}}, \bibinfo {author}
  {\bibfnamefont {G.}~\bibnamefont {Giriat}}, \bibinfo {author} {\bibfnamefont
  {D.}~\bibnamefont {Jaccard}}, \bibinfo {author} {\bibfnamefont
  {C.}~\bibnamefont {Berthod}}, \bibinfo {author} {\bibfnamefont {T.~K.}\
  \bibnamefont {Kim}}, \bibinfo {author} {\bibfnamefont {M.}~\bibnamefont
  {Hoesch}}, \bibinfo {author} {\bibfnamefont {E.~C.}\ \bibnamefont {Hunter}},
  \bibinfo {author} {\bibfnamefont {R.~S.}\ \bibnamefont {Perry}}, \bibinfo
  {author} {\bibfnamefont {A.}~\bibnamefont {Tamai}}, \ and\ \bibinfo {author}
  {\bibfnamefont {F.}~\bibnamefont {Baumberger}},\ }\href {\doibase
  10.1103/PhysRevLett.115.176402} {\bibfield  {journal} {\bibinfo  {journal}
  {Phys. Rev. Lett.}\ }\textbf {\bibinfo {volume} {115}},\ \bibinfo {pages}
  {176402} (\bibinfo {year} {2015})}\BibitemShut {NoStop}%
\bibitem [{\citenamefont {Zare}\ \emph {et~al.}(2017)\citenamefont {Zare},
  \citenamefont {Biderang},\ and\ \citenamefont {Akbari}}]{Biderang_PRB_2017}%
  \BibitemOpen
  \bibfield  {author} {\bibinfo {author} {\bibfnamefont {M.-H.}\ \bibnamefont
  {Zare}}, \bibinfo {author} {\bibfnamefont {M.}~\bibnamefont {Biderang}}, \
  and\ \bibinfo {author} {\bibfnamefont {A.}~\bibnamefont {Akbari}},\ }\href
  {\doibase 10.1103/PhysRevB.96.205156} {\bibfield  {journal} {\bibinfo
  {journal} {Phys. Rev. B}\ }\textbf {\bibinfo {volume} {96}},\ \bibinfo
  {pages} {205156} (\bibinfo {year} {2017})}\BibitemShut {NoStop}%
\bibitem [{\citenamefont {Nelson}\ \emph {et~al.}(2020)\citenamefont {Nelson},
  \citenamefont {Parzyck}, \citenamefont {Faeth}, \citenamefont {Kawasaki},
  \citenamefont {Schlom},\ and\ \citenamefont {Shen}}]{Nelson_Nat_Com_2020}%
  \BibitemOpen
  \bibfield  {author} {\bibinfo {author} {\bibfnamefont {J.}~\bibnamefont
  {Nelson}}, \bibinfo {author} {\bibfnamefont {C.}~\bibnamefont {Parzyck}},
  \bibinfo {author} {\bibfnamefont {B.}~\bibnamefont {Faeth}}, \bibinfo
  {author} {\bibfnamefont {J.}~\bibnamefont {Kawasaki}}, \bibinfo {author}
  {\bibfnamefont {D.}~\bibnamefont {Schlom}}, \ and\ \bibinfo {author}
  {\bibfnamefont {M.}~\bibnamefont {Shen}},\ }\href {\doibase
  10.1038/s41467-020-16425-z} {\bibfield  {journal} {\bibinfo  {journal}
  {Nature Communications}\ }\textbf {\bibinfo {volume} {11}} (\bibinfo {year}
  {2020}),\ 10.1038/s41467-020-16425-z}\BibitemShut {NoStop}%
\bibitem [{\citenamefont {Gor'kov}\ and\ \citenamefont
  {Rashba}(2001)}]{Grokov_PRL_2001}%
  \BibitemOpen
  \bibfield  {author} {\bibinfo {author} {\bibfnamefont {L.~P.}\ \bibnamefont
  {Gor'kov}}\ and\ \bibinfo {author} {\bibfnamefont {E.~I.}\ \bibnamefont
  {Rashba}},\ }\href {\doibase 10.1103/PhysRevLett.87.037004} {\bibfield
  {journal} {\bibinfo  {journal} {Phys. Rev. Lett.}\ }\textbf {\bibinfo
  {volume} {87}},\ \bibinfo {pages} {037004} (\bibinfo {year}
  {2001})}\BibitemShut {NoStop}%
\bibitem [{\citenamefont {Bauer}\ \emph {et~al.}(2004)\citenamefont {Bauer},
  \citenamefont {Hilscher}, \citenamefont {Michor}, \citenamefont {Paul},
  \citenamefont {Scheidt}, \citenamefont {Gribanov}, \citenamefont {Seropegin},
  \citenamefont {No{\"e}l}, \citenamefont {Sigrist},\ and\ \citenamefont
  {Rogl}}]{Sigrist_PRL_2004}%
  \BibitemOpen
  \bibfield  {author} {\bibinfo {author} {\bibfnamefont {E.}~\bibnamefont
  {Bauer}}, \bibinfo {author} {\bibfnamefont {G.}~\bibnamefont {Hilscher}},
  \bibinfo {author} {\bibfnamefont {H.}~\bibnamefont {Michor}}, \bibinfo
  {author} {\bibfnamefont {C.}~\bibnamefont {Paul}}, \bibinfo {author}
  {\bibfnamefont {E.~W.}\ \bibnamefont {Scheidt}}, \bibinfo {author}
  {\bibfnamefont {A.}~\bibnamefont {Gribanov}}, \bibinfo {author}
  {\bibfnamefont {Y.}~\bibnamefont {Seropegin}}, \bibinfo {author}
  {\bibfnamefont {H.}~\bibnamefont {No{\"e}l}}, \bibinfo {author}
  {\bibfnamefont {M.}~\bibnamefont {Sigrist}}, \ and\ \bibinfo {author}
  {\bibfnamefont {P.}~\bibnamefont {Rogl}},\ }\href {\doibase
  10.1103/PhysRevLett.92.027003} {\bibfield  {journal} {\bibinfo  {journal}
  {Phys. Rev. Lett.}\ }\textbf {\bibinfo {volume} {92}},\ \bibinfo {pages}
  {027003} (\bibinfo {year} {2004})}\BibitemShut {NoStop}%
\bibitem [{\citenamefont {Yogi}\ \emph {et~al.}(2004)\citenamefont {Yogi},
  \citenamefont {Kitaoka}, \citenamefont {Hashimoto}, \citenamefont {Yasuda},
  \citenamefont {Settai}, \citenamefont {Matsuda}, \citenamefont {Haga},
  \citenamefont {{\ifmmode \bar{O}\else {\=O}\fi{}nuki}}, \citenamefont
  {Rogl},\ and\ \citenamefont {Bauer}}]{Bauer_PRL_2004}%
  \BibitemOpen
  \bibfield  {author} {\bibinfo {author} {\bibfnamefont {M.}~\bibnamefont
  {Yogi}}, \bibinfo {author} {\bibfnamefont {Y.}~\bibnamefont {Kitaoka}},
  \bibinfo {author} {\bibfnamefont {S.}~\bibnamefont {Hashimoto}}, \bibinfo
  {author} {\bibfnamefont {T.}~\bibnamefont {Yasuda}}, \bibinfo {author}
  {\bibfnamefont {R.}~\bibnamefont {Settai}}, \bibinfo {author} {\bibfnamefont
  {T.~D.}\ \bibnamefont {Matsuda}}, \bibinfo {author} {\bibfnamefont
  {Y.}~\bibnamefont {Haga}}, \bibinfo {author} {\bibfnamefont {Y.}~\bibnamefont
  {{\ifmmode \bar{O}\else {\=O}\fi{}nuki}}}, \bibinfo {author} {\bibfnamefont
  {P.}~\bibnamefont {Rogl}}, \ and\ \bibinfo {author} {\bibfnamefont
  {E.}~\bibnamefont {Bauer}},\ }\href {\doibase 10.1103/PhysRevLett.93.027003}
  {\bibfield  {journal} {\bibinfo  {journal} {Phys. Rev. Lett.}\ }\textbf
  {\bibinfo {volume} {93}},\ \bibinfo {pages} {027003} (\bibinfo {year}
  {2004})}\BibitemShut {NoStop}%
\bibitem [{\citenamefont {Samokhin}\ \emph {et~al.}(2004)\citenamefont
  {Samokhin}, \citenamefont {Zijlstra},\ and\ \citenamefont
  {Bose}}]{Samokhin_PRB_2004}%
  \BibitemOpen
  \bibfield  {author} {\bibinfo {author} {\bibfnamefont {K.~V.}\ \bibnamefont
  {Samokhin}}, \bibinfo {author} {\bibfnamefont {E.~S.}\ \bibnamefont
  {Zijlstra}}, \ and\ \bibinfo {author} {\bibfnamefont {S.~K.}\ \bibnamefont
  {Bose}},\ }\href {\doibase 10.1103/PhysRevB.69.094514} {\bibfield  {journal}
  {\bibinfo  {journal} {Phys. Rev. B}\ }\textbf {\bibinfo {volume} {69}},\
  \bibinfo {pages} {094514} (\bibinfo {year} {2004})}\BibitemShut {NoStop}%
\bibitem [{\citenamefont {Frigeri}\ \emph {et~al.}(2004)\citenamefont
  {Frigeri}, \citenamefont {Agterberg}, \citenamefont {Koga},\ and\
  \citenamefont {Sigrist}}]{Frigeri_PRL_2004}%
  \BibitemOpen
  \bibfield  {author} {\bibinfo {author} {\bibfnamefont {P.~A.}\ \bibnamefont
  {Frigeri}}, \bibinfo {author} {\bibfnamefont {D.~F.}\ \bibnamefont
  {Agterberg}}, \bibinfo {author} {\bibfnamefont {A.}~\bibnamefont {Koga}}, \
  and\ \bibinfo {author} {\bibfnamefont {M.}~\bibnamefont {Sigrist}},\ }\href
  {\doibase 10.1103/PhysRevLett.92.097001} {\bibfield  {journal} {\bibinfo
  {journal} {Phys. Rev. Lett.}\ }\textbf {\bibinfo {volume} {92}},\ \bibinfo
  {pages} {097001} (\bibinfo {year} {2004})}\BibitemShut {NoStop}%
\bibitem [{\citenamefont {Fujimoto}(2007)}]{Fujimoto_Jpn_2007}%
  \BibitemOpen
  \bibfield  {author} {\bibinfo {author} {\bibfnamefont {S.}~\bibnamefont
  {Fujimoto}},\ }\href {\doibase 10.1143/JPSJ.76.051008} {\bibfield  {journal}
  {\bibinfo  {journal} {Journal of the Physical Society of Japan}\ }\textbf
  {\bibinfo {volume} {76}},\ \bibinfo {pages} {051008} (\bibinfo {year}
  {2007})}\BibitemShut {NoStop}%
\bibitem [{\citenamefont {Yanase}(2010)}]{Yanase_Jpn_2010}%
  \BibitemOpen
  \bibfield  {author} {\bibinfo {author} {\bibfnamefont {Y.}~\bibnamefont
  {Yanase}},\ }\href {\doibase 10.1143/JPSJ.79.084701} {\bibfield  {journal}
  {\bibinfo  {journal} {Journal of the Physical Society of Japan}\ }\textbf
  {\bibinfo {volume} {79}},\ \bibinfo {pages} {084701} (\bibinfo {year}
  {2010})}\BibitemShut {NoStop}%
\bibitem [{\citenamefont {Bauer}\ and\ \citenamefont
  {Sigrist}(2012)}]{Bauer_Sigrist_NCS_Book_2012}%
  \BibitemOpen
  \bibfield  {author} {\bibinfo {author} {\bibfnamefont {E.}~\bibnamefont
  {Bauer}}\ and\ \bibinfo {author} {\bibfnamefont {M.}~\bibnamefont
  {Sigrist}},\ }\href@noop {} {\emph {\bibinfo {title} {Non-Centrosymmetric
  Superconductors: Introduction and Overview}}}\ (\bibinfo  {publisher}
  {Springer Berlin Heidelberg},\ \bibinfo {year} {2012})\BibitemShut {NoStop}%
\bibitem [{\citenamefont {Biderang}\ \emph {et~al.}(2018)\citenamefont
  {Biderang}, \citenamefont {Yavari}, \citenamefont {Zare}, \citenamefont
  {Thalmeier},\ and\ \citenamefont {Akbari}}]{Biderang_PRB_2018}%
  \BibitemOpen
  \bibfield  {author} {\bibinfo {author} {\bibfnamefont {M.}~\bibnamefont
  {Biderang}}, \bibinfo {author} {\bibfnamefont {H.}~\bibnamefont {Yavari}},
  \bibinfo {author} {\bibfnamefont {M.-H.}\ \bibnamefont {Zare}}, \bibinfo
  {author} {\bibfnamefont {P.}~\bibnamefont {Thalmeier}}, \ and\ \bibinfo
  {author} {\bibfnamefont {A.}~\bibnamefont {Akbari}},\ }\href {\doibase
  10.1103/PhysRevB.98.014524} {\bibfield  {journal} {\bibinfo  {journal} {Phys.
  Rev. B}\ }\textbf {\bibinfo {volume} {98}},\ \bibinfo {pages} {014524}
  (\bibinfo {year} {2018})}\BibitemShut {NoStop}%
\bibitem [{\citenamefont {Greco}\ and\ \citenamefont
  {Schnyder}(2018)}]{Greco_PRL_2018}%
  \BibitemOpen
  \bibfield  {author} {\bibinfo {author} {\bibfnamefont {A.}~\bibnamefont
  {Greco}}\ and\ \bibinfo {author} {\bibfnamefont {A.~P.}\ \bibnamefont
  {Schnyder}},\ }\href {\doibase 10.1103/PhysRevLett.120.177002} {\bibfield
  {journal} {\bibinfo  {journal} {Phys. Rev. Lett.}\ }\textbf {\bibinfo
  {volume} {120}},\ \bibinfo {pages} {177002} (\bibinfo {year}
  {2018})}\BibitemShut {NoStop}%
\bibitem [{\citenamefont {Greco}\ \emph {et~al.}(2020)\citenamefont {Greco},
  \citenamefont {Bejas},\ and\ \citenamefont {Schnyder}}]{Greco_PRB_2020}%
  \BibitemOpen
  \bibfield  {author} {\bibinfo {author} {\bibfnamefont {A.}~\bibnamefont
  {Greco}}, \bibinfo {author} {\bibfnamefont {M.}~\bibnamefont {Bejas}}, \ and\
  \bibinfo {author} {\bibfnamefont {A.~P.}\ \bibnamefont {Schnyder}},\ }\href
  {\doibase 10.1103/PhysRevB.101.174420} {\bibfield  {journal} {\bibinfo
  {journal} {Phys. Rev. B}\ }\textbf {\bibinfo {volume} {101}},\ \bibinfo
  {pages} {174420} (\bibinfo {year} {2020})}\BibitemShut {NoStop}%
\bibitem [{\citenamefont {Qi}\ and\ \citenamefont
  {Zhang}(2011)}]{Qi_RevModPhys_2011}%
  \BibitemOpen
  \bibfield  {author} {\bibinfo {author} {\bibfnamefont {X.-L.}\ \bibnamefont
  {Qi}}\ and\ \bibinfo {author} {\bibfnamefont {S.-C.}\ \bibnamefont {Zhang}},\
  }\href {\doibase 10.1103/RevModPhys.83.1057} {\bibfield  {journal} {\bibinfo
  {journal} {Rev. Mod. Phys.}\ }\textbf {\bibinfo {volume} {83}},\ \bibinfo
  {pages} {1057} (\bibinfo {year} {2011})}\BibitemShut {NoStop}%
\bibitem [{\citenamefont {Beenakker}(2015)}]{Beenaker_RevModPhys_2015}%
  \BibitemOpen
  \bibfield  {author} {\bibinfo {author} {\bibfnamefont {C.~W.~J.}\
  \bibnamefont {Beenakker}},\ }\href {\doibase 10.1103/RevModPhys.87.1037}
  {\bibfield  {journal} {\bibinfo  {journal} {Rev. Mod. Phys.}\ }\textbf
  {\bibinfo {volume} {87}},\ \bibinfo {pages} {1037} (\bibinfo {year}
  {2015})}\BibitemShut {NoStop}%
\bibitem [{\citenamefont {Fischer}\ \emph {et~al.}(2011)\citenamefont
  {Fischer}, \citenamefont {Loder},\ and\ \citenamefont
  {Sigrist}}]{Fiscer_SIgrist_PRB_2011}%
  \BibitemOpen
  \bibfield  {author} {\bibinfo {author} {\bibfnamefont {M.~H.}\ \bibnamefont
  {Fischer}}, \bibinfo {author} {\bibfnamefont {F.}~\bibnamefont {Loder}}, \
  and\ \bibinfo {author} {\bibfnamefont {M.}~\bibnamefont {Sigrist}},\ }\href
  {\doibase 10.1103/PhysRevB.84.184533} {\bibfield  {journal} {\bibinfo
  {journal} {Phys. Rev. B}\ }\textbf {\bibinfo {volume} {84}},\ \bibinfo
  {pages} {184533} (\bibinfo {year} {2011})}\BibitemShut {NoStop}%
\bibitem [{\citenamefont {Sumita}\ \emph {et~al.}(2017)\citenamefont {Sumita},
  \citenamefont {Nomoto},\ and\ \citenamefont {Yanase}}]{Yanase_PRL_2017}%
  \BibitemOpen
  \bibfield  {author} {\bibinfo {author} {\bibfnamefont {S.}~\bibnamefont
  {Sumita}}, \bibinfo {author} {\bibfnamefont {T.}~\bibnamefont {Nomoto}}, \
  and\ \bibinfo {author} {\bibfnamefont {Y.}~\bibnamefont {Yanase}},\ }\href
  {\doibase 10.1103/PhysRevLett.119.027001} {\bibfield  {journal} {\bibinfo
  {journal} {Phys. Rev. Lett.}\ }\textbf {\bibinfo {volume} {119}},\ \bibinfo
  {pages} {027001} (\bibinfo {year} {2017})}\BibitemShut {NoStop}%
\bibitem [{\citenamefont {Ishizuka}\ and\ \citenamefont
  {Yanase}(2018)}]{Yanase_PRB_2018}%
  \BibitemOpen
  \bibfield  {author} {\bibinfo {author} {\bibfnamefont {J.}~\bibnamefont
  {Ishizuka}}\ and\ \bibinfo {author} {\bibfnamefont {Y.}~\bibnamefont
  {Yanase}},\ }\href {\doibase 10.1103/PhysRevB.98.224510} {\bibfield
  {journal} {\bibinfo  {journal} {Phys. Rev. B}\ }\textbf {\bibinfo {volume}
  {98}},\ \bibinfo {pages} {224510} (\bibinfo {year} {2018})}\BibitemShut
  {NoStop}%
\bibitem [{\citenamefont {Shaked}\ \emph {et~al.}(2000)\citenamefont {Shaked},
  \citenamefont {Jorgensen}, \citenamefont {Chmaissem}, \citenamefont {Ikeda},\
  and\ \citenamefont {Maeno}}]{Shaked_2000}%
  \BibitemOpen
  \bibfield  {author} {\bibinfo {author} {\bibfnamefont {H.}~\bibnamefont
  {Shaked}}, \bibinfo {author} {\bibfnamefont {J.}~\bibnamefont {Jorgensen}},
  \bibinfo {author} {\bibfnamefont {O.}~\bibnamefont {Chmaissem}}, \bibinfo
  {author} {\bibfnamefont {S.}~\bibnamefont {Ikeda}}, \ and\ \bibinfo {author}
  {\bibfnamefont {Y.}~\bibnamefont {Maeno}},\ }\href {\doibase
  10.1006/jssc.2000.8796} {\bibfield  {journal} {\bibinfo  {journal} {Journal
  of Solid State Chemistry}\ }\textbf {\bibinfo {volume} {154}},\ \bibinfo
  {pages} {361} (\bibinfo {year} {2000})}\BibitemShut {NoStop}%
\bibitem [{\citenamefont {Subramanian}\ \emph {et~al.}(1994)\citenamefont
  {Subramanian}, \citenamefont {Crawford}, \citenamefont {Harlow},
  \citenamefont {Ami}, \citenamefont {Fernandez-Baca}, \citenamefont {Wang},\
  and\ \citenamefont {Johnston}}]{SUBRAMANIAN_1994}%
  \BibitemOpen
  \bibfield  {author} {\bibinfo {author} {\bibfnamefont {M.}~\bibnamefont
  {Subramanian}}, \bibinfo {author} {\bibfnamefont {M.}~\bibnamefont
  {Crawford}}, \bibinfo {author} {\bibfnamefont {R.}~\bibnamefont {Harlow}},
  \bibinfo {author} {\bibfnamefont {T.}~\bibnamefont {Ami}}, \bibinfo {author}
  {\bibfnamefont {J.}~\bibnamefont {Fernandez-Baca}}, \bibinfo {author}
  {\bibfnamefont {Z.}~\bibnamefont {Wang}}, \ and\ \bibinfo {author}
  {\bibfnamefont {D.}~\bibnamefont {Johnston}},\ }\href {\doibase
  10.1016/0921-4534(94)91596-2} {\bibfield  {journal} {\bibinfo  {journal}
  {Physica C: Superconductivity}\ }\textbf {\bibinfo {volume} {235-240}},\
  \bibinfo {pages} {743} (\bibinfo {year} {1994})}\BibitemShut {NoStop}%
\bibitem [{\citenamefont {Huang}\ \emph {et~al.}(1994)\citenamefont {Huang},
  \citenamefont {Soubeyroux}, \citenamefont {Chmaissem}, \citenamefont {Sora},
  \citenamefont {Santoro}, \citenamefont {Cava}, \citenamefont {Krajewski},\
  and\ \citenamefont {Peck}}]{Huang_JSSC_1994}%
  \BibitemOpen
  \bibfield  {author} {\bibinfo {author} {\bibfnamefont {Q.}~\bibnamefont
  {Huang}}, \bibinfo {author} {\bibfnamefont {J.}~\bibnamefont {Soubeyroux}},
  \bibinfo {author} {\bibfnamefont {O.}~\bibnamefont {Chmaissem}}, \bibinfo
  {author} {\bibfnamefont {I.}~\bibnamefont {Sora}}, \bibinfo {author}
  {\bibfnamefont {A.}~\bibnamefont {Santoro}}, \bibinfo {author} {\bibfnamefont
  {R.}~\bibnamefont {Cava}}, \bibinfo {author} {\bibfnamefont {J.}~\bibnamefont
  {Krajewski}}, \ and\ \bibinfo {author} {\bibfnamefont {W.}~\bibnamefont
  {Peck}},\ }\href {\doibase 10.1006/jssc.1994.1316} {\bibfield  {journal}
  {\bibinfo  {journal} {Journal of Solid State Chemistry}\ }\textbf {\bibinfo
  {volume} {112}},\ \bibinfo {pages} {355} (\bibinfo {year}
  {1994})}\BibitemShut {NoStop}%
\bibitem [{\citenamefont {Crawford}\ \emph {et~al.}(1994)\citenamefont
  {Crawford}, \citenamefont {Subramanian}, \citenamefont {Harlow},
  \citenamefont {Fernandez-Baca}, \citenamefont {Wang},\ and\ \citenamefont
  {Johnston}}]{Huang_PRB_1994}%
  \BibitemOpen
  \bibfield  {author} {\bibinfo {author} {\bibfnamefont {M.~K.}\ \bibnamefont
  {Crawford}}, \bibinfo {author} {\bibfnamefont {M.~A.}\ \bibnamefont
  {Subramanian}}, \bibinfo {author} {\bibfnamefont {R.~L.}\ \bibnamefont
  {Harlow}}, \bibinfo {author} {\bibfnamefont {J.~A.}\ \bibnamefont
  {Fernandez-Baca}}, \bibinfo {author} {\bibfnamefont {Z.~R.}\ \bibnamefont
  {Wang}}, \ and\ \bibinfo {author} {\bibfnamefont {D.~C.}\ \bibnamefont
  {Johnston}},\ }\href {\doibase 10.1103/PhysRevB.49.9198} {\bibfield
  {journal} {\bibinfo  {journal} {Phys. Rev. B}\ }\textbf {\bibinfo {volume}
  {49}},\ \bibinfo {pages} {9198} (\bibinfo {year} {1994})}\BibitemShut
  {NoStop}%
\bibitem [{\citenamefont {Ye}\ \emph {et~al.}(2015)\citenamefont {Ye},
  \citenamefont {Wang}, \citenamefont {Hoffmann}, \citenamefont {Wang},
  \citenamefont {Chi}, \citenamefont {Matsuda}, \citenamefont {Chakoumakos},
  \citenamefont {Fernandez-Baca},\ and\ \citenamefont {Cao}}]{Ye_PRB_2015}%
  \BibitemOpen
  \bibfield  {author} {\bibinfo {author} {\bibfnamefont {F.}~\bibnamefont
  {Ye}}, \bibinfo {author} {\bibfnamefont {X.}~\bibnamefont {Wang}}, \bibinfo
  {author} {\bibfnamefont {C.}~\bibnamefont {Hoffmann}}, \bibinfo {author}
  {\bibfnamefont {J.}~\bibnamefont {Wang}}, \bibinfo {author} {\bibfnamefont
  {S.}~\bibnamefont {Chi}}, \bibinfo {author} {\bibfnamefont {M.}~\bibnamefont
  {Matsuda}}, \bibinfo {author} {\bibfnamefont {B.~C.}\ \bibnamefont
  {Chakoumakos}}, \bibinfo {author} {\bibfnamefont {J.~A.}\ \bibnamefont
  {Fernandez-Baca}}, \ and\ \bibinfo {author} {\bibfnamefont {G.}~\bibnamefont
  {Cao}},\ }\href {\doibase 10.1103/PhysRevB.92.201112} {\bibfield  {journal}
  {\bibinfo  {journal} {Phys. Rev. B}\ }\textbf {\bibinfo {volume} {92}},\
  \bibinfo {pages} {201112} (\bibinfo {year} {2015})}\BibitemShut {NoStop}%
\bibitem [{\citenamefont {Torchinsky}\ \emph
  {et~al.}(2015{\natexlab{a}})\citenamefont {Torchinsky}, \citenamefont {Chu},
  \citenamefont {Zhao}, \citenamefont {Perkins}, \citenamefont {Sizyuk},
  \citenamefont {Qi}, \citenamefont {Cao},\ and\ \citenamefont
  {Hsieh}}]{Torchinsky_PRL_2015}%
  \BibitemOpen
  \bibfield  {author} {\bibinfo {author} {\bibfnamefont {D.~H.}\ \bibnamefont
  {Torchinsky}}, \bibinfo {author} {\bibfnamefont {H.}~\bibnamefont {Chu}},
  \bibinfo {author} {\bibfnamefont {L.}~\bibnamefont {Zhao}}, \bibinfo {author}
  {\bibfnamefont {N.~B.}\ \bibnamefont {Perkins}}, \bibinfo {author}
  {\bibfnamefont {Y.}~\bibnamefont {Sizyuk}}, \bibinfo {author} {\bibfnamefont
  {T.}~\bibnamefont {Qi}}, \bibinfo {author} {\bibfnamefont {G.}~\bibnamefont
  {Cao}}, \ and\ \bibinfo {author} {\bibfnamefont {D.}~\bibnamefont {Hsieh}},\
  }\href {\doibase 10.1103/PhysRevLett.114.096404} {\bibfield  {journal}
  {\bibinfo  {journal} {Phys. Rev. Lett.}\ }\textbf {\bibinfo {volume} {114}},\
  \bibinfo {pages} {096404} (\bibinfo {year} {2015}{\natexlab{a}})}\BibitemShut
  {NoStop}%
\bibitem [{\citenamefont {Gretarsson}\ \emph {et~al.}(2016)\citenamefont
  {Gretarsson}, \citenamefont {Sung}, \citenamefont {Porras}, \citenamefont
  {Bertinshaw}, \citenamefont {Dietl}, \citenamefont {Bruin}, \citenamefont
  {Bangura}, \citenamefont {Kim}, \citenamefont {Dinnebier}, \citenamefont
  {Kim}, \citenamefont {Al-Zein}, \citenamefont {{Moretti Sala}}, \citenamefont
  {Krisch}, \citenamefont {{Le Tacon}}, \citenamefont {Keimer},\ and\
  \citenamefont {Kim}}]{BJ_Kim_PRL_2016}%
  \BibitemOpen
  \bibfield  {author} {\bibinfo {author} {\bibfnamefont {H.}~\bibnamefont
  {Gretarsson}}, \bibinfo {author} {\bibfnamefont {N.~H.}\ \bibnamefont
  {Sung}}, \bibinfo {author} {\bibfnamefont {J.}~\bibnamefont {Porras}},
  \bibinfo {author} {\bibfnamefont {J.}~\bibnamefont {Bertinshaw}}, \bibinfo
  {author} {\bibfnamefont {C.}~\bibnamefont {Dietl}}, \bibinfo {author}
  {\bibfnamefont {J.~A.~N.}\ \bibnamefont {Bruin}}, \bibinfo {author}
  {\bibfnamefont {A.~F.}\ \bibnamefont {Bangura}}, \bibinfo {author}
  {\bibfnamefont {Y.~K.}\ \bibnamefont {Kim}}, \bibinfo {author} {\bibfnamefont
  {R.}~\bibnamefont {Dinnebier}}, \bibinfo {author} {\bibfnamefont
  {J.}~\bibnamefont {Kim}}, \bibinfo {author} {\bibfnamefont {A.}~\bibnamefont
  {Al-Zein}}, \bibinfo {author} {\bibfnamefont {M.}~\bibnamefont {{Moretti
  Sala}}}, \bibinfo {author} {\bibfnamefont {M.}~\bibnamefont {Krisch}},
  \bibinfo {author} {\bibfnamefont {M.}~\bibnamefont {{Le Tacon}}}, \bibinfo
  {author} {\bibfnamefont {B.}~\bibnamefont {Keimer}}, \ and\ \bibinfo {author}
  {\bibfnamefont {B.~J.}\ \bibnamefont {Kim}},\ }\href {\doibase
  10.1103/PhysRevLett.117.107001} {\bibfield  {journal} {\bibinfo  {journal}
  {Phys. Rev. Lett.}\ }\textbf {\bibinfo {volume} {117}},\ \bibinfo {pages}
  {107001} (\bibinfo {year} {2016})}\BibitemShut {NoStop}%
\bibitem [{\citenamefont {Liu}\ \emph {et~al.}(2016)\citenamefont {Liu},
  \citenamefont {Dean}, \citenamefont {Meng}, \citenamefont {Upton},
  \citenamefont {Qi}, \citenamefont {Gog}, \citenamefont {Cao}, \citenamefont
  {Lin}, \citenamefont {Meyers}, \citenamefont {Ding}, \citenamefont {Cao},\
  and\ \citenamefont {Hill}}]{Hil_PRB_2016}%
  \BibitemOpen
  \bibfield  {author} {\bibinfo {author} {\bibfnamefont {X.}~\bibnamefont
  {Liu}}, \bibinfo {author} {\bibfnamefont {M.~P.~M.}\ \bibnamefont {Dean}},
  \bibinfo {author} {\bibfnamefont {Z.~Y.}\ \bibnamefont {Meng}}, \bibinfo
  {author} {\bibfnamefont {M.~H.}\ \bibnamefont {Upton}}, \bibinfo {author}
  {\bibfnamefont {T.}~\bibnamefont {Qi}}, \bibinfo {author} {\bibfnamefont
  {T.}~\bibnamefont {Gog}}, \bibinfo {author} {\bibfnamefont {Y.}~\bibnamefont
  {Cao}}, \bibinfo {author} {\bibfnamefont {J.~Q.}\ \bibnamefont {Lin}},
  \bibinfo {author} {\bibfnamefont {D.}~\bibnamefont {Meyers}}, \bibinfo
  {author} {\bibfnamefont {H.}~\bibnamefont {Ding}}, \bibinfo {author}
  {\bibfnamefont {G.}~\bibnamefont {Cao}}, \ and\ \bibinfo {author}
  {\bibfnamefont {J.~P.}\ \bibnamefont {Hill}},\ }\href {\doibase
  10.1103/PhysRevB.93.241102} {\bibfield  {journal} {\bibinfo  {journal} {Phys.
  Rev. B}\ }\textbf {\bibinfo {volume} {93}},\ \bibinfo {pages} {241102}
  (\bibinfo {year} {2016})}\BibitemShut {NoStop}%
\bibitem [{\citenamefont {Pincini}\ \emph {et~al.}(2017)\citenamefont
  {Pincini}, \citenamefont {Vale}, \citenamefont {Donnerer}, \citenamefont
  {de~la Torre}, \citenamefont {Hunter}, \citenamefont {Perry}, \citenamefont
  {{Moretti Sala}}, \citenamefont {Baumberger},\ and\ \citenamefont
  {McMorrow}}]{Pincini_PRB_2017}%
  \BibitemOpen
  \bibfield  {author} {\bibinfo {author} {\bibfnamefont {D.}~\bibnamefont
  {Pincini}}, \bibinfo {author} {\bibfnamefont {J.~G.}\ \bibnamefont {Vale}},
  \bibinfo {author} {\bibfnamefont {C.}~\bibnamefont {Donnerer}}, \bibinfo
  {author} {\bibfnamefont {A.}~\bibnamefont {de~la Torre}}, \bibinfo {author}
  {\bibfnamefont {E.~C.}\ \bibnamefont {Hunter}}, \bibinfo {author}
  {\bibfnamefont {R.}~\bibnamefont {Perry}}, \bibinfo {author} {\bibfnamefont
  {M.}~\bibnamefont {{Moretti Sala}}}, \bibinfo {author} {\bibfnamefont
  {F.}~\bibnamefont {Baumberger}}, \ and\ \bibinfo {author} {\bibfnamefont
  {D.~F.}\ \bibnamefont {McMorrow}},\ }\href {\doibase
  10.1103/PhysRevB.96.075162} {\bibfield  {journal} {\bibinfo  {journal} {Phys.
  Rev. B}\ }\textbf {\bibinfo {volume} {96}},\ \bibinfo {pages} {075162}
  (\bibinfo {year} {2017})}\BibitemShut {NoStop}%
\bibitem [{\citenamefont {{Kivelson}}\ \emph {et~al.}(1998)\citenamefont
  {{Kivelson}}, \citenamefont {{Fradkin}},\ and\ \citenamefont
  {{Emery}}}]{Kivelson_Nature_1998}%
  \BibitemOpen
  \bibfield  {author} {\bibinfo {author} {\bibfnamefont {S.~A.}\ \bibnamefont
  {{Kivelson}}}, \bibinfo {author} {\bibfnamefont {E.}~\bibnamefont
  {{Fradkin}}}, \ and\ \bibinfo {author} {\bibfnamefont {V.~J.}\ \bibnamefont
  {{Emery}}},\ }\href {\doibase 10.1038/31177} {\bibfield  {journal} {\bibinfo
  {journal} {\nat}\ }\textbf {\bibinfo {volume} {393}},\ \bibinfo {pages} {550}
  (\bibinfo {year} {1998})}\BibitemShut {NoStop}%
\bibitem [{\citenamefont {Moon}\ \emph {et~al.}(2009)\citenamefont {Moon},
  \citenamefont {Jin}, \citenamefont {Choi}, \citenamefont {Lee}, \citenamefont
  {Seo}, \citenamefont {Yu}, \citenamefont {Cao}, \citenamefont {Noh},\ and\
  \citenamefont {Lee}}]{Moon_PRB_2009}%
  \BibitemOpen
  \bibfield  {author} {\bibinfo {author} {\bibfnamefont {S.~J.}\ \bibnamefont
  {Moon}}, \bibinfo {author} {\bibfnamefont {H.}~\bibnamefont {Jin}}, \bibinfo
  {author} {\bibfnamefont {W.~S.}\ \bibnamefont {Choi}}, \bibinfo {author}
  {\bibfnamefont {J.~S.}\ \bibnamefont {Lee}}, \bibinfo {author} {\bibfnamefont
  {S.~S.~A.}\ \bibnamefont {Seo}}, \bibinfo {author} {\bibfnamefont
  {J.}~\bibnamefont {Yu}}, \bibinfo {author} {\bibfnamefont {G.}~\bibnamefont
  {Cao}}, \bibinfo {author} {\bibfnamefont {T.~W.}\ \bibnamefont {Noh}}, \ and\
  \bibinfo {author} {\bibfnamefont {Y.~S.}\ \bibnamefont {Lee}},\ }\href
  {\doibase 10.1103/PhysRevB.80.195110} {\bibfield  {journal} {\bibinfo
  {journal} {Phys. Rev. B}\ }\textbf {\bibinfo {volume} {80}},\ \bibinfo
  {pages} {195110} (\bibinfo {year} {2009})}\BibitemShut {NoStop}%
\bibitem [{\citenamefont {Ye}\ \emph {et~al.}(2013)\citenamefont {Ye},
  \citenamefont {Chi}, \citenamefont {Chakoumakos}, \citenamefont
  {Fernandez-Baca}, \citenamefont {Qi},\ and\ \citenamefont
  {Cao}}]{Ye_PRB_2013}%
  \BibitemOpen
  \bibfield  {author} {\bibinfo {author} {\bibfnamefont {F.}~\bibnamefont
  {Ye}}, \bibinfo {author} {\bibfnamefont {S.}~\bibnamefont {Chi}}, \bibinfo
  {author} {\bibfnamefont {B.~C.}\ \bibnamefont {Chakoumakos}}, \bibinfo
  {author} {\bibfnamefont {J.~A.}\ \bibnamefont {Fernandez-Baca}}, \bibinfo
  {author} {\bibfnamefont {T.}~\bibnamefont {Qi}}, \ and\ \bibinfo {author}
  {\bibfnamefont {G.}~\bibnamefont {Cao}},\ }\href {\doibase
  10.1103/PhysRevB.87.140406} {\bibfield  {journal} {\bibinfo  {journal} {Phys.
  Rev. B}\ }\textbf {\bibinfo {volume} {87}},\ \bibinfo {pages} {140406}
  (\bibinfo {year} {2013})}\BibitemShut {NoStop}%
\bibitem [{\citenamefont {Torchinsky}\ \emph
  {et~al.}(2015{\natexlab{b}})\citenamefont {Torchinsky}, \citenamefont {Chu},
  \citenamefont {Zhao}, \citenamefont {Perkins}, \citenamefont {Sizyuk},
  \citenamefont {Qi}, \citenamefont {Cao},\ and\ \citenamefont
  {Hsieh}}]{Hsieh_PRL_2015}%
  \BibitemOpen
  \bibfield  {author} {\bibinfo {author} {\bibfnamefont {D.~H.}\ \bibnamefont
  {Torchinsky}}, \bibinfo {author} {\bibfnamefont {H.}~\bibnamefont {Chu}},
  \bibinfo {author} {\bibfnamefont {L.}~\bibnamefont {Zhao}}, \bibinfo {author}
  {\bibfnamefont {N.~B.}\ \bibnamefont {Perkins}}, \bibinfo {author}
  {\bibfnamefont {Y.}~\bibnamefont {Sizyuk}}, \bibinfo {author} {\bibfnamefont
  {T.}~\bibnamefont {Qi}}, \bibinfo {author} {\bibfnamefont {G.}~\bibnamefont
  {Cao}}, \ and\ \bibinfo {author} {\bibfnamefont {D.}~\bibnamefont {Hsieh}},\
  }\href {\doibase 10.1103/PhysRevLett.114.096404} {\bibfield  {journal}
  {\bibinfo  {journal} {Phys. Rev. Lett.}\ }\textbf {\bibinfo {volume} {114}},\
  \bibinfo {pages} {096404} (\bibinfo {year} {2015}{\natexlab{b}})}\BibitemShut
  {NoStop}%
\bibitem [{\citenamefont {Kim}\ \emph {et~al.}(2008)\citenamefont {Kim},
  \citenamefont {Jin}, \citenamefont {Moon}, \citenamefont {Kim}, \citenamefont
  {Park}, \citenamefont {Leem}, \citenamefont {Yu}, \citenamefont {Noh},
  \citenamefont {Kim}, \citenamefont {Oh}, \citenamefont {Park}, \citenamefont
  {Durairaj}, \citenamefont {Cao},\ and\ \citenamefont
  {Rotenberg}}]{BJKim_PRL_2008}%
  \BibitemOpen
  \bibfield  {author} {\bibinfo {author} {\bibfnamefont {B.~J.}\ \bibnamefont
  {Kim}}, \bibinfo {author} {\bibfnamefont {H.}~\bibnamefont {Jin}}, \bibinfo
  {author} {\bibfnamefont {S.~J.}\ \bibnamefont {Moon}}, \bibinfo {author}
  {\bibfnamefont {J.-Y.}\ \bibnamefont {Kim}}, \bibinfo {author} {\bibfnamefont
  {B.-G.}\ \bibnamefont {Park}}, \bibinfo {author} {\bibfnamefont {C.~S.}\
  \bibnamefont {Leem}}, \bibinfo {author} {\bibfnamefont {J.}~\bibnamefont
  {Yu}}, \bibinfo {author} {\bibfnamefont {T.~W.}\ \bibnamefont {Noh}},
  \bibinfo {author} {\bibfnamefont {C.}~\bibnamefont {Kim}}, \bibinfo {author}
  {\bibfnamefont {S.-J.}\ \bibnamefont {Oh}}, \bibinfo {author} {\bibfnamefont
  {J.-H.}\ \bibnamefont {Park}}, \bibinfo {author} {\bibfnamefont
  {V.}~\bibnamefont {Durairaj}}, \bibinfo {author} {\bibfnamefont
  {G.}~\bibnamefont {Cao}}, \ and\ \bibinfo {author} {\bibfnamefont
  {E.}~\bibnamefont {Rotenberg}},\ }\href {\doibase
  10.1103/PhysRevLett.101.076402} {\bibfield  {journal} {\bibinfo  {journal}
  {Phys. Rev. Lett.}\ }\textbf {\bibinfo {volume} {101}},\ \bibinfo {pages}
  {076402} (\bibinfo {year} {2008})}\BibitemShut {NoStop}%
\bibitem [{\citenamefont {Wang}\ and\ \citenamefont
  {Senthil}(2011{\natexlab{b}})}]{Senthil_prl_2011}%
  \BibitemOpen
  \bibfield  {author} {\bibinfo {author} {\bibfnamefont {F.}~\bibnamefont
  {Wang}}\ and\ \bibinfo {author} {\bibfnamefont {T.}~\bibnamefont {Senthil}},\
  }\href {\doibase 10.1103/PhysRevLett.106.136402} {\bibfield  {journal}
  {\bibinfo  {journal} {Phys. Rev. Lett.}\ }\textbf {\bibinfo {volume} {106}},\
  \bibinfo {pages} {136402} (\bibinfo {year} {2011}{\natexlab{b}})}\BibitemShut
  {NoStop}%
\bibitem [{\citenamefont {Jin}\ \emph {et~al.}(2009)\citenamefont {Jin},
  \citenamefont {Jeong}, \citenamefont {Ozaki},\ and\ \citenamefont
  {Yu}}]{Jaejun_PRB_2009}%
  \BibitemOpen
  \bibfield  {author} {\bibinfo {author} {\bibfnamefont {H.}~\bibnamefont
  {Jin}}, \bibinfo {author} {\bibfnamefont {H.}~\bibnamefont {Jeong}}, \bibinfo
  {author} {\bibfnamefont {T.}~\bibnamefont {Ozaki}}, \ and\ \bibinfo {author}
  {\bibfnamefont {J.}~\bibnamefont {Yu}},\ }\href {\doibase
  10.1103/PhysRevB.80.075112} {\bibfield  {journal} {\bibinfo  {journal} {Phys.
  Rev. B}\ }\textbf {\bibinfo {volume} {80}},\ \bibinfo {pages} {075112}
  (\bibinfo {year} {2009})}\BibitemShut {NoStop}%
\bibitem [{\citenamefont {Young}\ and\ \citenamefont
  {Kane}(2015)}]{Kane_PRL_2015}%
  \BibitemOpen
  \bibfield  {author} {\bibinfo {author} {\bibfnamefont {S.~M.}\ \bibnamefont
  {Young}}\ and\ \bibinfo {author} {\bibfnamefont {C.~L.}\ \bibnamefont
  {Kane}},\ }\href {\doibase 10.1103/PhysRevLett.115.126803} {\bibfield
  {journal} {\bibinfo  {journal} {Phys. Rev. Lett.}\ }\textbf {\bibinfo
  {volume} {115}},\ \bibinfo {pages} {126803} (\bibinfo {year}
  {2015})}\BibitemShut {NoStop}%
\bibitem [{\citenamefont {Moutenet}\ \emph {et~al.}(2018)\citenamefont
  {Moutenet}, \citenamefont {Georges},\ and\ \citenamefont
  {Ferrero}}]{Ferrero_PRB_2018}%
  \BibitemOpen
  \bibfield  {author} {\bibinfo {author} {\bibfnamefont {A.}~\bibnamefont
  {Moutenet}}, \bibinfo {author} {\bibfnamefont {A.}~\bibnamefont {Georges}}, \
  and\ \bibinfo {author} {\bibfnamefont {M.}~\bibnamefont {Ferrero}},\ }\href
  {\doibase 10.1103/PhysRevB.97.155109} {\bibfield  {journal} {\bibinfo
  {journal} {Phys. Rev. B}\ }\textbf {\bibinfo {volume} {97}},\ \bibinfo
  {pages} {155109} (\bibinfo {year} {2018})}\BibitemShut {NoStop}%
\bibitem [{\citenamefont {Lindquist}\ and\ \citenamefont
  {Kee}(2019)}]{Kee_PRB_2019}%
  \BibitemOpen
  \bibfield  {author} {\bibinfo {author} {\bibfnamefont {A.~W.}\ \bibnamefont
  {Lindquist}}\ and\ \bibinfo {author} {\bibfnamefont {H.-Y.}\ \bibnamefont
  {Kee}},\ }\href {\doibase 10.1103/PhysRevB.100.054512} {\bibfield  {journal}
  {\bibinfo  {journal} {Phys. Rev. B}\ }\textbf {\bibinfo {volume} {100}},\
  \bibinfo {pages} {054512} (\bibinfo {year} {2019})}\BibitemShut {NoStop}%
\bibitem [{\citenamefont {Kung}\ \emph {et~al.}(2015)\citenamefont {Kung},
  \citenamefont {Nowadnick}, \citenamefont {Jia}, \citenamefont {Johnston},
  \citenamefont {Moritz}, \citenamefont {Scalettar},\ and\ \citenamefont
  {Devereaux}}]{Devereaux_PRB_2015}%
  \BibitemOpen
  \bibfield  {author} {\bibinfo {author} {\bibfnamefont {Y.~F.}\ \bibnamefont
  {Kung}}, \bibinfo {author} {\bibfnamefont {E.~A.}\ \bibnamefont {Nowadnick}},
  \bibinfo {author} {\bibfnamefont {C.~J.}\ \bibnamefont {Jia}}, \bibinfo
  {author} {\bibfnamefont {S.}~\bibnamefont {Johnston}}, \bibinfo {author}
  {\bibfnamefont {B.}~\bibnamefont {Moritz}}, \bibinfo {author} {\bibfnamefont
  {R.~T.}\ \bibnamefont {Scalettar}}, \ and\ \bibinfo {author} {\bibfnamefont
  {T.~P.}\ \bibnamefont {Devereaux}},\ }\href {\doibase
  10.1103/PhysRevB.92.195108} {\bibfield  {journal} {\bibinfo  {journal} {Phys.
  Rev. B}\ }\textbf {\bibinfo {volume} {92}},\ \bibinfo {pages} {195108}
  (\bibinfo {year} {2015})}\BibitemShut {NoStop}%
\bibitem [{\citenamefont {Cobo}\ \emph {et~al.}(2016)\citenamefont {Cobo},
  \citenamefont {Ahn}, \citenamefont {Eremin},\ and\ \citenamefont
  {Akbari}}]{Cobo_PRB_2016}%
  \BibitemOpen
  \bibfield  {author} {\bibinfo {author} {\bibfnamefont {S.}~\bibnamefont
  {Cobo}}, \bibinfo {author} {\bibfnamefont {F.}~\bibnamefont {Ahn}}, \bibinfo
  {author} {\bibfnamefont {I.}~\bibnamefont {Eremin}}, \ and\ \bibinfo {author}
  {\bibfnamefont {A.}~\bibnamefont {Akbari}},\ }\href {\doibase
  10.1103/PhysRevB.94.224507} {\bibfield  {journal} {\bibinfo  {journal} {Phys.
  Rev. B}\ }\textbf {\bibinfo {volume} {94}},\ \bibinfo {pages} {224507}
  (\bibinfo {year} {2016})}\BibitemShut {NoStop}%
\bibitem [{\citenamefont {Ghadimi}\ \emph {et~al.}(2019)\citenamefont
  {Ghadimi}, \citenamefont {Kargarian},\ and\ \citenamefont
  {Jafari}}]{Ghadimi_PRB_2019}%
  \BibitemOpen
  \bibfield  {author} {\bibinfo {author} {\bibfnamefont {R.}~\bibnamefont
  {Ghadimi}}, \bibinfo {author} {\bibfnamefont {M.}~\bibnamefont {Kargarian}},
  \ and\ \bibinfo {author} {\bibfnamefont {S.~A.}\ \bibnamefont {Jafari}},\
  }\href {\doibase 10.1103/PhysRevB.99.115122} {\bibfield  {journal} {\bibinfo
  {journal} {Phys. Rev. B}\ }\textbf {\bibinfo {volume} {99}},\ \bibinfo
  {pages} {115122} (\bibinfo {year} {2019})}\BibitemShut {NoStop}%
\bibitem [{\citenamefont {{Pereira}}\ \emph {et~al.}(2007)\citenamefont
  {{Pereira}}, \citenamefont {{Sirker}}, \citenamefont {{Caux}}, \citenamefont
  {{Hagemans}}, \citenamefont {{Maillet}}, \citenamefont {{White}},\ and\
  \citenamefont {{Affleck}}}]{Jesko_IOP_2007}%
  \BibitemOpen
  \bibfield  {author} {\bibinfo {author} {\bibfnamefont {R.~G.}\ \bibnamefont
  {{Pereira}}}, \bibinfo {author} {\bibfnamefont {J.}~\bibnamefont {{Sirker}}},
  \bibinfo {author} {\bibfnamefont {J.~S.}\ \bibnamefont {{Caux}}}, \bibinfo
  {author} {\bibfnamefont {R.}~\bibnamefont {{Hagemans}}}, \bibinfo {author}
  {\bibfnamefont {J.~M.}\ \bibnamefont {{Maillet}}}, \bibinfo {author}
  {\bibfnamefont {S.~R.}\ \bibnamefont {{White}}}, \ and\ \bibinfo {author}
  {\bibfnamefont {I.}~\bibnamefont {{Affleck}}},\ }\href {\doibase
  10.1088/1742-5468/2007/08/P08022} {\bibfield  {journal} {\bibinfo  {journal}
  {Journal of Statistical Mechanics: Theory and Experiment}\ }\textbf {\bibinfo
  {volume} {2007}},\ \bibinfo {pages} {08022} (\bibinfo {year}
  {2007})}\BibitemShut {NoStop}%
\bibitem [{\citenamefont {Akbari}\ and\ \citenamefont
  {Khaliullin}(2014)}]{Alireza_PRB_2014}%
  \BibitemOpen
  \bibfield  {author} {\bibinfo {author} {\bibfnamefont {A.}~\bibnamefont
  {Akbari}}\ and\ \bibinfo {author} {\bibfnamefont {G.}~\bibnamefont
  {Khaliullin}},\ }\href {\doibase 10.1103/PhysRevB.90.035137} {\bibfield
  {journal} {\bibinfo  {journal} {Phys. Rev. B}\ }\textbf {\bibinfo {volume}
  {90}},\ \bibinfo {pages} {035137} (\bibinfo {year} {2014})}\BibitemShut
  {NoStop}%
\bibitem [{\citenamefont {Bertinshaw}\ \emph {et~al.}(2020)\citenamefont
  {Bertinshaw}, \citenamefont {Kim}, \citenamefont {Porras}, \citenamefont
  {Ueda}, \citenamefont {Sung}, \citenamefont {Efimenko}, \citenamefont
  {Bombardi}, \citenamefont {Kim}, \citenamefont {Keimer},\ and\ \citenamefont
  {Kim}}]{BJ_Kim_PRB_2020}%
  \BibitemOpen
  \bibfield  {author} {\bibinfo {author} {\bibfnamefont {J.}~\bibnamefont
  {Bertinshaw}}, \bibinfo {author} {\bibfnamefont {J.~K.}\ \bibnamefont {Kim}},
  \bibinfo {author} {\bibfnamefont {J.}~\bibnamefont {Porras}}, \bibinfo
  {author} {\bibfnamefont {K.}~\bibnamefont {Ueda}}, \bibinfo {author}
  {\bibfnamefont {N.~H.}\ \bibnamefont {Sung}}, \bibinfo {author}
  {\bibfnamefont {A.}~\bibnamefont {Efimenko}}, \bibinfo {author}
  {\bibfnamefont {A.}~\bibnamefont {Bombardi}}, \bibinfo {author}
  {\bibfnamefont {J.}~\bibnamefont {Kim}}, \bibinfo {author} {\bibfnamefont
  {B.}~\bibnamefont {Keimer}}, \ and\ \bibinfo {author} {\bibfnamefont {B.~J.}\
  \bibnamefont {Kim}},\ }\href {\doibase 10.1103/PhysRevB.101.094428}
  {\bibfield  {journal} {\bibinfo  {journal} {Phys. Rev. B}\ }\textbf {\bibinfo
  {volume} {101}},\ \bibinfo {pages} {094428} (\bibinfo {year}
  {2020})}\BibitemShut {NoStop}%
\end{thebibliography}%

\end{document}